\documentclass[12pt]{article}
\usepackage{amsmath,amsthm,amssymb}
\usepackage{float}
\usepackage{bm}
\usepackage{graphicx}
\usepackage{subfigure}
\usepackage{enumerate}
\usepackage{hyperref}
\usepackage{natbib}
\usepackage{url} % not crucial - just used below for the URL 
\usepackage[dvipsnames]{xcolor}
\usepackage{algorithm}
\usepackage{algpseudocode}

\hypersetup{
    colorlinks=true,
    linkcolor=Maroon,   
    urlcolor=NavyBlue,
    citecolor=Green,
}

\newcommand{\blind}{1}

\newtheorem{theorem}{Theorem}
\newtheorem{lemma}{Lemma}

\theoremstyle{definition}

\newtheorem{definition}{Definition}

\newtheorem{assumption}{Assumption}

\newcommand{\argmin}{\operatorname*{\arg\min}}
\newcommand{\E}{\operatorname{\mathbb{E}}} % expectation
\newcommand{\bU}{{\mathbf U}}

\newcommand{\bX}{{\mathbf X}}

\newcommand{\bx}{{\mathbf x}}

\newcommand{\bbeta}  {\boldsymbol{\beta}}

\newcommand{\bgamma}{\boldsymbol{\gamma}}

\newcommand{\bPhi} {\boldsymbol{\Phi}}
\newcommand{\bPsi} {\boldsymbol{\Psi}}

\newcommand{\pa}{{\text{PA}}}
\newcommand{\ch}{{\text{CH}}}
\newcommand{\nd}{{\text{ND}}}
\begin{document}

\def\spacingset#1{\renewcommand{\baselinestretch}%
{#1}\small\normalsize} \spacingset{1}

%%%%%%%%%%%%%%%%%%%%%%%%%%%%%%%%%%%%%%%%%%%%%%%%%%%%%%%%%%%%%%%%%%%%%%%%%%%%%%

\newcommand{\Title}{Learning a directed acyclic graph with additive heteroscedastic errors}

\if1\blind
{
  \title{\bf \Title}
    \author{Xintao Xia\thanks{Center for Data Science, Zhejiang University. Email: \url{xintaox@zju.edu.cn}.}
    \and 
    Li Chen\thanks{
    School of Statistics, University of Minnesota. Email: \url{chen7019@umn.edu}.}
    \and
    Yue Hu\thanks{
    Center for Data Science, Zhejiang University. Email: \url{hy1015@zju.edu.cn}.}
    \and 
    Chunlin Li\thanks{Department of Statistics, University of Virginia. Email: \url{chunlin@virginia.edu}.} $^{,}$\thanks{Department of Statistics, Iowa State University.}}
    \date{}
  \maketitle
} \fi

\if0\blind
{
  \bigskip
  \bigskip
  \bigskip
  \title{\bf \Title}
  \maketitle
  \medskip
} \fi

\bigskip
\begin{abstract}
This paper studies causal discovery for a directed acyclic graph under a structural equation model with additive heteroscedastic errors. We first establish new identifiability results for location-scale noise models, showing that heteroscedasticity can be leveraged to recover causal directions. Based on these insights, we propose a novel iterative procedure, Residual Simultaneous Quantile Estimation (RESQUE), where each iteration consists of a residual-construction stage and a composite quantile regression stage, enabling recursive identification of sink nodes via the invariance of conditional scale coefficients across quantiles. We then establish its theoretical guarantees for recovering topological order and graph structure, even when the number of variables diverges with the sample size. Simulation studies and application to benchmark datasets show that RESQUE performs favorably compared with existing methods, especially when causal information is partly encoded in the variance component. These results highlight exploiting structured variance signals for causal discovery and provide a principled framework for multivariate causal discovery beyond mean-based modeling.
\end{abstract}

\noindent%

%\newpage
\spacingset{1.65} % DON'T change the spacing!
% 1.65: 28 lines per page: JRSSB (a4paper)
% 1.65: 26 lines per page: JASA (usletter)
% 1.7: 25 lines per page (usletter)

\section{Introduction}
\label{sec:intro}

Establishing causal relations enables formal reasoning about interventions and counterfactuals, and has been an important topic in modern science and engineering, with applications in biology \citep{sachs2005causal, maathuis2010predicting}, econometrics \citep{wold1954causality}, psychometrics \citep{bentler1988causal}, industrial manufacturing \citep{yang2024hierarchical}, dynamical systems \cite{li2020causal}, artificial intelligence \citep{scholkopf2021toward}, and other domains. 
While randomized experiments are the gold standard for causality, they are often unethical or infeasible. 
Causal discovery aims to learn causal relations among variables from observational data \citep{glymour2019review}, moving beyond statistical associations to determine the direction of effects. It is useful for formulating data-driven conjectures in scientific studies and is viewed as a crucial step towards building automated learning and reasoning systems \citep{scholkopf2021toward}. 

Formally, causal relations among variables are often represented using a directed acyclic graph (DAG) \citep{spirtes2000causation, pearl2009causality}, a directed graph with no directed cycles, where nodes represent variables of interest, directed edges encode direct causal effects, and the absence of an edge indicates the lack of a direct causal relation. The acyclicity constraint rules out feedback loops, ensuring that the associated data-generating process is well defined \citep{peters2014causal}. The goal of causal discovery is to reconstruct the underlying DAG.

With solely observational data, determining causal directions is impossible without additional identifiability assumptions on the data-generating process, which is often described by a structural equation model (SEM). Such identifiability assumptions often impose restrictions on the function class or error distributions in the SEM. From a modeling perspective, these conditions reflect a researcher's inductive bias toward simplicity, modularity, and interpretability in the underlying causal mechanisms. 
For example, LINGAM assumes linear structural equations while requiring the errors to be non-Gaussian and homoscedastic \citep{shimizu2006linear}; SEMs with homoscedastic Gaussian additive errors require the functions to be nonlinear and smooth \citep{peters2014causal}. 
Recently, due to the prevalence of heteroscedasticity observed in data, causal models with heteroscedastic errors have received increasing attention \citep{immer2023identifiability,xu2022inferring,tran2024robust,yin2024effective,lin2025skewness}; however, they mainly focus on bivariate cases while multivariate causal discovery remains understudied. The challenge is two-fold. 
First, existing identifiability results \citep{immer2023identifiability, strobl2023identifying,lin2025skewness} are not applicable to bounded variables. It is still unclear under what assumptions the DAG is identifiable when there are bounded and/or unbounded (continuous) variables. 
Second, likelihood-based methods are prone to parametric misspecification in heteroscedastic models \cite{sun2023cause,schultheiss2023pitfalls}, leading to inconsistent estimation. 
Meanwhile, independence testing based methods, such as those based on HSIC \citep{zhang2023statistical} or mutual information \citep{strobl2023identifying}, often require sample splitting and struggle in high dimensions \citep{berrevoets2025differentiable}, as they do not fully exploit the structured signals in variance components. 

In this work, we study causal discovery under an SEM with heteroscedastic errors and address the above questions. Specifically, we contribute to the following aspects:
\begin{enumerate}
    \item We establish new identifiability results for DAGs with additive heteroscedastic errors. These results unify the identifiability theory for both bounded and unbounded continuous variables in multivariate causal discovery, accommodating practically important yet largely overlooked scenarios. In particular, we find that causal discovery among bounded variables requires much milder conditions than for unbounded variables, thereby complementing the existing theory.
    
    \item We propose a novel iterative estimation procedure, Residual Simultaneous Quantile Estimation (RESQUE), with each iteration consisting of a residual construction step and a composite quantile regression step, integrated with shrinkage techniques to exploit heteroscedastic errors. The proposed approach recursively identifies sink nodes via their parallel conditional quantiles. 
    Thus, it examines error independence rather than optimizing a likelihood function, circumventing the reliance on parametric specification. It exploits the structured, sparse signals in variance components by transforming independence testing to feature selection. In this way, RESQUE avoids sample splitting and many high-dimensional (nonparametric) independence tests, leading to improved performance in causal discovery. 
    
    \item We analyze the theoretical properties of RESQUE and prove its consistency for order and graph estimation. The simulations and real-data benchmarks demonstrate the strong performance of RESQUE, consistent with our motivation and theoretical analysis.
\end{enumerate}

The central message is two-fold: bounded variables help with the identifiability of heteroscedastic SEMs, and structured causal signals in heteroscedastic patterns should be exploited for multivariate causal discovery.

\paragraph{Related work.} 

Causal discovery has been extensively studied; see \cite{heinze2018causal,glymour2019review,vowels2022d} for comprehensive reviews. Broadly, there are four distinct but interconnected approaches to causal discovery from solely observational data. 
(a) Constraint-based methods \cite{spirtes1991algorithm, harris2013pc, spirtes2001anytime} test conditional independence to infer the underlying causal graph up to its Markov equivalence classes.
(b) Asymmetry-based methods 
exploit the algebraic asymmetry implied by SEMs to determine the causal order. Prominent examples include order-based approaches \cite{peters2014causal, zhao2022learning} and independent component analysis approaches \cite{shimizu2006linear, zhang2009identifiability}.
(c) Score-based methods estimate the causal graph by optimizing a certain criterion function over a discrete \cite{chickering2002optimal} or a continuous search space \cite{zheng2018dags,li2020likelihood}. 
(d) Hybrid methods \cite{tsamardinos2006max,peter2014cam} seek to combine a constraint-based approach with an asymmetry- or score-based approach. Our work belongs to the family of asymmetry-based methods.

Among the above approaches, (b)--(d) often introduce an SEM, also called functional causal model, to describe the data-generating process, where the underlying DAG is induced by the SEM. Thus, the identifiability of the SEM has been a central topic in causal discovery \cite{shimizu2006linear,zhang2010distinguishing,zhang2009identifiability,hoyer2009nonlinear,peters2010identifying,peters2014causal,blobaum2018cause,park2020identifiability}. 
Notably, \cite{shimizu2006linear} studies multivariate linear SEMs with non-Gaussian errors; \cite{zhang2009identifiability} comprehensively characterizes bivariate post-nonlinear causal models; 
\cite{peters2014causal} thoroughly discusses the identifiability of multivariate additive noise models (ANMs); and \cite{immer2023identifiability, strobl2023identifying, lin2025skewness} offer identifiability results for heteroscedastic SEMs. 
Nonetheless, most work has focused exclusively on continuous variables with full (unbounded) support.
Exceptions include \cite{peters2010identifying}, which discusses ANMs with discrete variables, and \cite{blobaum2018cause}, which studies ANMs with compactly supported variables and smooth monotonic functional relations.
Our work adds to the prior literature by offering a unifying identification for multivariate heteroscedastic SEMs with bounded and unbounded continuous variables.

\paragraph{Organization.}
The rest of the paper is organized as follows. Section \ref{sec:model} introduces the additive heteroscedastic causal model and presents the identifiability results. Section \ref{sec:resque} describes the proposed finite-sample estimation method and Section \ref{sec:theory} establishes its theoretical properties. Section \ref{sec:numerical-study} examines the method through simulation studies and real data benchmarks. Finally, Section \ref{sec:discussion} concludes the paper with a discussion of extensions and future directions.

\section{Heteroscedastic causal model and identifiability}
\label{sec:model}

We first introduce the basic terminology of DAGs. A DAG is defined as a graph in which every edge is directed, and the graph does not contain any loops or cycles. Let $G=(V,E)$ be a DAG associated with a random vector $\bX\in\mathbb{R}^p$. The node set of $V=\{X_1,\dots,X_p\}$ indexes the components of $\bX$, and a directed edge $(i,j)\in E$ encodes a direct causal relationship from $X_i$ to $X_j$. If $(i,j)\in E$, then $X_i$ is called a parent of $X_j$, and $X_j$ is called a child of $X_i$. The set of parents of $X_j$ is denoted by $\pa^G(j)$, and the set of children of $X_i$ is denoted by $\ch^G(i)$. If there exists a directed path from $X_j$ to $X_k$ in $G$, then $X_j$ is called an ancestor of $X_k$. The set of all ancestors of $X_j$, including $X_j$, is denoted by $\text{AN}^G(j)$, while the set of strict ancestors is denoted by $\text{an}^G(j)= \text{AN}^G(j)\setminus\{j\}$. The set of all non-descendants of $X_j$, including $X_j$, is denoted by $\nd^G(j)$.

We consider the following data-generating procedure: \textit{independent and identically distributed} (i.i.d.) realizations of the random vector $\{X_j\}_{j=1}^{p}$ are generated according to a \textit{structural equation model} (SEM) associated with a DAG \citep{pearl2009causality}:
\begin{equation*}
X_j=h_j(\boldsymbol{X}_{\pa^G(j)},\varepsilon_j)\quad\text{for }j=1,\dots,p,
\end{equation*}
where $\{\varepsilon_1,\dots,\varepsilon_p\}$ are mutually independent random errors, and $h_j(\cdot):\mathbb{R}^{|\pa^{G}(j)|+1}\to\mathbb{R}$ are unknown functions. Lacking additional assumptions on the SEM, the underlying DAG $G$ is identifiable only up to its Markov equivalence class. Given a DAG $G=(V,E)$, a permutation $\pi$, defined as a bijection from the node set $[p]$ to $[p]$, is considered a \textit{topological order (causal order)} of $G$ if, for any $X_i,X_j\in V$, such that $X_i$ is an ancestor of $X_j$, we have $\pi(i)<\pi(j)$. See Definition~\ref{def:correct_topo} for a formal statement. A permutation $\pi$, represented by its ordered sequence $(\pi(1),\dots,\pi(p))$, does not in general uniquely determine a DAG. For example, the causal order $(1,2,3)$ is compatible with multiple DAGs, including one with directed edges $X_1\to X_2\to X_3$ and another with only a single edge $X_1\to X_2$. Conversely, a DAG may admit multiple causal orders. Nevertheless, a topological order remains highly informative. We study its identifiability as a first step toward identifying the DAG.

\begin{definition}
\label{def:correct_topo}
We say that the algorithm recovers a topological order $\pi$ over the DAG if 
\begin{equation*}
    \pi(i)<\pi(j)\quad\text{implies}\quad j\notin \text{an}(i,G)\quad\text{for all}\quad i,j\in V.
\end{equation*}
\end{definition}

To fully exploit the information contained in the error distribution, we consider a general \textit{location-scale noise model} (LSNM) allowing the noise to depend on parent variables:
\begin{equation}
\label{eq:lsnm}
    X_j=f_j(\bX_{\pa^{G}(j)})+g_j(\bX_{\pa^{G}(j)})\varepsilon_j,
\end{equation}
where $f_j(\cdot):\mathbb{R}^{|\pa^{G}(j)|}\to\mathbb{R}$ and $g_j(\cdot):\mathbb{R}^{|\pa^{G}(j)|}\to\mathbb{R}^+$ are unknown functions. We first study the bivariate case and then extend the analysis to the general multivariate setting.

\begin{assumption}[Bivariate Identifiability]
    \label{ass:biv_lsnm_identifiable}
     Suppose the two-variable SEM \eqref{eq:lsnm} given by: $X_i = \varepsilon_i$ and $X_j = f(X_i) + g(X_i)\varepsilon_j$, where $f,g$ are not constant functions and the joint distribution of $(X_i,X_j)$ is absolutely continuous. For any reverse SEM, $X_i = h(X_j) + k(X_j)\tilde{\varepsilon}_i$ and $X_j = \tilde{\varepsilon}_j$, with densities $p_{X_j}(\cdot)$ and $p_{\tilde{\varepsilon}_i|X_j}(\cdot|\cdot)$, one of the following conditions must hold:
    \begin{enumerate}
        \item Let $\nu_1(\cdot):=\log p_{X_j}(\cdot)$, $\nu_2(\cdot,\cdot):=\log p_{\tilde{\varepsilon}_i|X_j}(\cdot|\cdot)$, and $f,g,h,k$ be twice-differentiable. The following equation fails to hold for all $ x_i,x_j$ satisfying $ p_{X_j}(x_j)>0$, $p_{\tilde{\varepsilon}_i|X_j}(x_i|x_j) > 0$ and $ H(x_i,x_j) \neq 0$, where $H(x_i,x_j)=g(x_i)f'(x_i)+g'(x_i)[x_j-f(x_i)]$:
        \begin{equation*}
            \begin{aligned}
                0=\nu_1^{\prime \prime}(x_j)+\frac{g^{\prime}(x_i)}{H(x_i, x_j)} \nu_1^{\prime}(x_j)+\bigg(\frac{\partial^2}{\partial x_j^2} +\frac{g(x_i)}{H(x_i, x_j)} \frac{\partial^2}{\partial x_j \partial x_i} +\frac{g^{\prime}(x_i)}{H(x_i, x_j)} \frac{\partial}{\partial x_j}\bigg) \nu_2(x_i,x_j).
            \end{aligned}
        \end{equation*}
        \item Let $\nu_3(\cdot,\cdot):=p_{X_i|X_j}(\cdot|\cdot)$, and $f,g,h,k$ be differentiable. The error distributions in \eqref{eq:lsnm} have continuous support, and $X_i$ has bounded support. Define $s(x) := \operatorname{ess}\sup( X_i \mid X_j = x ) - \operatorname{ess}\inf ( X_i \mid X_j = x )$ and $\ell(x) := \operatorname{ess}\inf(X_i/s(x) | X_j = x )$, where $\operatorname{ess}\sup(X_i|X_j)$ (respectively, $\operatorname{ess}\inf(X_i|X_j)$) denotes the smallest upper (largest lower) bound of the conditional distribution that holds almost surely. Either the set $\{ x_i\in \mathbb{R} : p_{X_i\mid X_j}(x_i \mid x_j) > 0 \}$ is disconnected for some $x_j$ with $p_{X_j}(x_j)>0$, or the following equation fail to hold for $z=x_i/s(x)-\ell(x)$ whenever $p_{X_j}(x_j)>0$, $p_{X_i|X_j}(x_i|x_j) > 0$:
        \begin{equation*}
0= s'(x)\nu_3(z,x)+ s(x)\Big[\partial_z \nu_3(z,x)\{s'(x)(z+\ell(x))+s(x)\ell'(x)\}+ \partial_x \nu_3(z,x)\Big].
\end{equation*}
    \end{enumerate}
\end{assumption}

\begin{theorem}
\label{thm:identifiability_biv_lsnm}
    Suppose the joint distribution of $(X_1, X_2)$ satisfies Assumption \ref{ass:biv_lsnm_identifiable} with underlying graph $G$. Then $G$ is identifiable from the joint distribution of $(X_1, X_2)$.
\end{theorem}
Theorem~\ref{thm:identifiability_biv_lsnm} establishes identifiability in the bivariate setting without requiring unbounded support, thereby relaxing a key assumption commonly imposed in the literature. The first condition is adopted from \cite{immer2023identifiability}, while the second leverages bounded domains to obtain identifiability in a broader class of models. In addition, due to space constraints, we establish identifiability via the boundary property in the Appendix.

\begin{theorem}[Identification of DAG]
\label{thm:identifiability_lsnm}
    Let $\boldsymbol{X}\in\mathbb{R}^p$ be generated according to the location–scale noise structural equation model in \eqref{eq:lsnm} with underlying DAG $G$ and the joint distribution of $\bX$ is absolutely continuous. Assume that for every $j \in [p]$, every $i \in \pa^{G}(j)$, and every set $\mathcal{S} \subseteq [p]$ with $\pa^{G}(j) \backslash \{i\} \subseteq \mathcal{S} \subseteq \nd^{G}(j) \backslash \{i,j\}$. Define $\mathcal{S}_1 = \pa^{G}(j) \backslash \{i\}$ and $\mathcal{S}_2 = \mathcal{S} \backslash \mathcal{S}_1$. There exists a realization $\bx_{\mathcal{S}}$ satisfying $p_{\bX_\mathcal{S}}(\bx_{\mathcal{S}})>0$, where $p_{\bX_\mathcal{S}}(\cdot)$ denotes the marginal density of $\bX_{\mathcal{S}}$, and the following conditions hold. Define the pseudo-variables $X_j^*:=f_j(X^*_i,\bx_{\mathcal{S}_1})+g_j(X^*_i,\bx_{\mathcal{S}_1})\varepsilon_j$ and $X_i^* \sim \mathcal{L}(X_i\mid \bX_{\mathcal{S}}=\bx_{\mathcal{S}})$. Then the pair $(X_i^*,X_j^*)$ satisfies Assumption \ref{ass:biv_lsnm_identifiable}. Then $G$ is identifiable from the joint distribution of $\bX$.
\end{theorem}

Theorem~\ref{thm:identifiability_lsnm} establishes identifiability of the DAG under a multivariate location–scale noise model. For each parent–child pair, it requires the existence of a conditioning under which the resulting conditional random variables (pseudo-variables) satisfy the bivariate identifiability assumptions. These conditions are local, requiring only a single realization with positive density. The result mirrors and extends the additive noise model identifiability results of \citep{peters2014identifiability}, providing a principled foundation for causal discovery in multivariate nonlinear models with heteroskedastic noise.

\section{Residual simultaneous quantile estimation}
\label{sec:resque}

In this section, we develop a causal discovery procedure based on conditional quantile functions. We first identify the topological order and then recover the causal graph.

\subsection{Recovering topological order}

In this section, we introduce a method that leverages comparisons of conditional quantile functions to iteratively identify sink nodes and thereby recover a topological ordering of the variables. Suppose that we observe $n$ \textit{independent and identically distributed} (i.i.d.) samples $\{\bX_i=(X_{i,1},\ldots,X_{i,p})^{\top}\}_{i=1}^{n}$ generated according to the model in \eqref{eq:lsnm},
\begin{equation*}
X_{i,j}=f_j(\bX_{i,\pa^{G}(j)})+g_j(\bX_{i,\pa^{G}(j)})\varepsilon_{i,j}\quad\text{for}\quad i=1,\dots,n; \ j=1,\dots,p.
\end{equation*}
For each $j=1,\ldots,p$, we approximate $f_j(\bX_{\pa^{G}(j)})$ and $\log g_j(\bX_{\pa^{G}(j)})$ via linear combinations of the feature maps $\bPhi(\bX_{\pa^{G}(j)})$ and $\bPsi(\bX_{\pa^{G}(j)})$, respectively. Here, $\bPhi(\cdot)$ and $\bPsi(\cdot)$ represent feature transformations, such as polynomial basis expansions of $\bX_{\pa^{G}(j)}$. To simplify notation, for any index set $A\subset[p]$ and integer $c_d\in\mathbb{N}$, we write $\Phi(\bX_A)$ and $\Psi(\bX_A)$ for feature maps that send $\bX_A\in\mathbb{R}^{|A|}$ to $\mathbb{R}^{c_d\times |A|}$. We assume that these feature maps are monotone in the sense that, for any $A\subset B\subset[p]$, $\text{span}(\Phi(\bX_{A}))\subset\text{span}(\Phi(\bX_{B}))$ and $\text{span}(\Psi(\bX_{A}))\subset\text{span}(\Psi(\bX_{B}))$ for $A\subset B\subset[p]$. It remains to consider the model:
\begin{equation}
\label{eq:gam-s}
    X_{j}=\bPhi(\bX_{\pa^{G}(j)})^{\top}\bbeta^*_j+r^{(f)}_{j}+\exp\{\bPsi(\bX_{\pa^{G}(j)})^{\top}\bgamma^*_j+r^{(g)}_{j}\}\varepsilon_{j},
\end{equation}
where $\bbeta_j^{*}$ and $\bgamma_j^{*}$ are best-approximation parameters, to be defined below, and $r_j^{(f)},r_j^{(g)}$ represent the approximation errors for $f_j(\bX_{\pa^{G}(j)})$ and $\log g_j(\bX_{\pa^{G}(j)})$, respectively. 

As a first step, we obtain the residuals by fitting a least squares regression model. We consider $p$ regression problems of the form regressing $X_j$ on $\bPhi(\bX_{-j})$ for $j=1,\ldots,p$. For each $j$, we define the empirical loss function $\mathcal{L}_{n,j}(\cdot):\mathbb{R}^{c_d(p-1)}\to\mathbb{R}$ by
\begin{equation}
\mathcal{L}_{n,j}(\bbeta)=\frac{1}{n}\sum_{i=1}^{n}\{X_{i,j}-\bPhi(\bX_{i,-j})^{\top}\bbeta\}^2/2+\lambda_1\|\bbeta\|_1
\label{eq:first-stage}
\end{equation}
where $\lambda_1>0$ is a tuning parameter, and $\bX_{i,-j}$ denotes $\bX_i$ with its $j$th component removed. The corresponding population parameter $\bbeta_j^{*}$ in \eqref{eq:gam-s} is defined as $\bbeta_j^{*}=\argmin_{\bbeta}\mathbb{E}[(X_{j}-\bPhi(\bX_{-j})^{\top}\bbeta)^2]$. Let $\widehat{\bbeta}_j=\arg\min_{\bbeta}\mathcal{L}_{n,j}(\bbeta)$ denote the corresponding sample estimator. We then define the estimated residuals as $\widehat{Y}_{i,j}:=\log|X_{i,j}-\bPhi(\bX_{i,-j})^{\top}\widehat{\bbeta}_j|$ for all $i,j$.

Let $0<\tau_1<\dots<\tau_K<1$ be $K$ quantiles of interest. We next estimate the conditional quantile functions $Q_{\tau_k}(\widehat{Y}_j\mid\bX_{-j})$ for $k\in\{1,\dots,K\}$. Since its introduction by \cite{koenker1978regression}, quantile regression has become a standard approach for estimating conditional quantile functions. Note that the conditional $\tau$th quantile function satisfies the following population estimating equation:
\begin{equation*}
(\bgamma_{j,k},b_{j,k})=\operatorname*{argmin}_{\bgamma,b}\E[\rho_{\tau_k}(Y_j-\bPsi(\bX_{-j})^{\top}\bgamma-b)],
\end{equation*}
where $Y_j:=\log|X_{j}-r_{j}^{(f)}-\bPhi(\bX_{-j})^{\top}\bbeta^*|$ and $\rho_{\tau}(u)=u\{\tau-1(u\leq 0)\}$ is the check function. When $X_j$ is a sink node, the coefficient $\bgamma_{j,k}$ is invariant across quantile levels and coincides with $\bgamma_{j}^*$ on the components associated with $\bPsi(\bX_{\pa^{G}(j)})$, while all remaining components are zero. We next consider a penalized composite quantile regression model to identify the sink nodes. For simplicity, we assume that the distribution of $\varepsilon_j$ is continuous, so that the quantile-specific intercepts $b_{j,k}$ are unique defined. Given the estimated residuals $\widehat{Y}_{i,j}$, we formulate the following penalized quantile regression problem:
\begin{align*}
&\{(\widehat{\bgamma}_{j,k},\widehat{b}_{j,k})\}_{k=1}^{K}\\
=&
\operatorname*{argmin}_{(\bgamma_k,b_k)_{k=1}^{K}}\sum_{k=1}^{K}\{\frac{1}{n}\sum_{i=1}^{n}\rho_{\tau_k}(\widehat{Y}_{i,j}-\bPsi(\bX_{i,-j})^{\top}\bgamma_k-b_k)+\lambda_2\|\bgamma_k\|_1\}+\sum_{k=1}^{K}P_{\lambda_3}(\|\bgamma_k-\bgamma_{k+1}\|_2),
\end{align*}
where $P_{\lambda_3}(\cdot)$ is a penalty function to be specified later, $\lambda_2$ and $\lambda_3$ are tuning parameters, and we adopt the convention $\bgamma_{K+1}=\bgamma_1$. A node is identified as a sink node if all estimated coefficient vectors $\widehat{\bgamma}_k$ for $1\leq k\leq K$are shrunk to the same value. We then remove all identified sink nodes from $[p]$ and apply the proposed procedure to the remaining nodes. This process is repeated until all nodes are identified. This iterative identification of sink nodes induces a topological ordering of the variables in reverse: nodes identified earlier appear later in the order. By successively removing sink nodes, we recover the topological order from sinks to sources (see Figure~\ref{fig:topo-order}).

We next address the computational aspects of the proposed procedure. The penalized optimization problem in the first-stage mean regression can be solved efficiently via \textit{proximal gradient descent}. For the second-stage penalized quantile regression, we adopt a \text{general folded concave penalty} \citep{fan2014strong}, which satisfies the following properties:
\begin{enumerate}
    \item $P_{\lambda}(t)$ is increasing and concave in $t\in[0,\infty)$ with $P_{\lambda}(0)=0$;
    \item $P_{\lambda}(t)$ is differentiable in $t\in[0,\infty)$ with $P_{\lambda}^{\prime}(+0)\geq a_1\lambda$;
    \item $P_{\lambda}^{\prime}(t)\geq a_1\lambda$ for $t\in(0,a_2\lambda]$;
    \item $P_{\lambda}^{\prime}(t)=0$ for $t\in[a\lambda,\infty)$ with pre-specified constant $a>a_2$.
\end{enumerate}
The above class of penalties includes SCAD \citep{fan2001variable} and MCP \citep{zhang2010nearly} as special cases. The resulting optimization problem can be efficiently solved via the local linear approximation (LLA) algorithm. Let $\{(\widehat{\bgamma}_{j,k}^{(0)}, \widehat{b}_{j,k}^{(0)})\}_{k=1}^{K}$ denote the solution to the penalized composite quantile regression problem
\begin{equation*}
\begin{split}
(\widehat{\bgamma}_{j,k}^{(0)},\widehat{b}_{j,k}^{(0)})_{k=1}^{K}=\operatorname*{argmin}_{(\bgamma_{k},b_{k})_{k=1}^{K}}&\sum_{k=1}^{K}\bigg\{\frac{1}{n}\sum_{i=1}^{n}\rho_{\tau_k}(\widehat{Y}_{i,j}-\bPsi(\bX_{i,-j})^{\top}\bgamma_{k}-b_{k})+\lambda_2\|\bgamma_{k}\|_1\bigg\},
\end{split}
\end{equation*}
which yields a convex optimization problem that can be solved efficiently using linear programming \citep{li20081}. The corresponding LLA-based optimizer is given by
\begin{equation}
\begin{split}
\label{eq:second-stage}
(\widehat{\bgamma}_{j,k},\widehat{b}_{j,k})_{k=1}^{K}=\operatorname*{argmin}_{(\bgamma_{k},b_{k})_{k=1}^{K}}&\sum_{k=1}^{K}\bigg\{\frac{1}{n}\sum_{i=1}^{n}\rho_{\tau_k}(\widehat{Y}_{i,j}-\bPsi(\bX_{i,-j})^{\top}\bgamma_{k}-b_{k})+\lambda_2\|\bgamma_{j,k}\|_1\bigg\}\\
&+\sum_{k=1}^{K}P_{\lambda_3}^{\prime}(\|\widehat{\bgamma}_{j,k}^{(0)}-\widehat{\bgamma}_{j,k+1}^{(0)}\|_2)\|\bgamma_{j,k}-\bgamma_{j,k+1}\|_2,
\end{split}
\end{equation}
where $P_{\lambda_3}^{\prime}(\cdot)$ denotes the derivative of $P_{\lambda_3}(\cdot)$. In practice, for a given $\lambda_3$, the estimated parameters $\widehat{\bgamma}_{j,k}$ for $j=1,\dots,p$ may not be exactly shrunk to the same value due to finite-sample variability. In such cases, we propose selecting the node that minimizes the discrepancy measure $\sum_{k=1}^{K}\|\widehat{\bgamma}_k-\widehat{\bgamma}_{k+1}\|_2$ as the sink node, and then proceeding with the iterative identification procedure.

\subsection{Recovering graph structure}
\label{subsec:iden_graph}

The proposed method can be further extended to recover the underlying graph structure. When $X_j$ is identified as a sink node, under the additive location–scale model \eqref{eq:lsnm}, $X_j=f_j(\bX_{\pa^{G}(j)})+g_j(\bX_{\pa^{G}(j)})\varepsilon_j$, there exists a directed edge from node $l$ to node $j$ if and only if either $f_j(\bX_{\pa^{G}(j)})$ or $g_j(\bX_{\pa^{G}(j)})$ is nonconstant as $X_l$ varies. With basis expansions, consider the working model $X_{j}=\bPhi(\bX_{-j})^{\top}\bbeta_j+\exp\{\bPsi(\bX_{-j})^{\top}\bgamma^*_j\}\varepsilon_{j}$. Under this representation, the above condition can be reformulated as a \textit{variable–selection} problem. Specifically, there is a directed edge from node $l$ to node $j$ if and only if there exists an index $q>0$ such that either (i) $[\bbeta_j]_q\neq 0$ and $[\bPhi(\bX_{-j})]_q$ is not a constant function of $X_l$, or (ii) $[\bgamma_{j,1}]_q\neq 0$ and $[\bPsi(\bX_{-j})]_q$ is not a constant function of $X_l$, where $[\cdot]_q$ denotes the $q$th component of a vector.

Building on the estimation procedures in \eqref{eq:first-stage} and \eqref{eq:second-stage}, we develop two approaches for graph recovery. We first adopt general folded concave penalties together with the local linear approximation (LLA) algorithm to achieve selection consistency. Given the preliminary estimates $\widehat{\bbeta}_j$ and $\widehat{\bgamma}_{j,1}$, we define the following one-step LLA estimators:
\begin{equation}
\begin{split}
\label{eq:lla-select}
&\widehat{\bbeta}_j^{c}=\arg\min_{\bbeta}\frac{1}{n}\sum_{i=1}^{n}\{X_{i,j}-\bPhi(\bX_{i,-j})^{\top}\bbeta\}^2/2+\sum_l P_{\lambda_1^c}^{\prime}(|\widehat{\beta}_{j,l}|)|\beta_l|,\\
&\widehat{\bgamma}_{j,1}^{c}=\arg\min_{\bgamma,b}\frac{1}{n}\sum_{i=1}^{n}\rho_{\tau_1}(\widehat{Y}_{i,j}-\bPsi(\bX_{i,-j})^{\top}\bgamma-b)+\sum_l P_{\lambda_2^c}^{\prime}(|\widehat{\gamma}_{j,1,l}|)|\gamma_l|,   
\end{split}
\end{equation}
where $\lambda_1^c,\lambda_2^c$ are penalties, and the summation over $l$ ranges over all components. Since all $\widehat{\gamma}_{j,k}$ are shrunk to the same value, we use $\widehat{\bgamma}_{j,1}$ for illustration. As established by \cite{fan2014strong}, under suitable regularity conditions the estimators $\widehat{\bbeta}_j^{c},\widehat{\bgamma}_{j,1}^{c}$ are \textit{sign-consistent} \citep{zhao2006model}, which in turn implies selection consistency. We also consider a second, computationally simpler approach based on hard thresholding. Specifically, the basis functions $[\bPhi(\bX_{-j})]_q$ and $[\bPsi(\bX_{-j})]_q$ are selected whenever $|[\bPhi(\bX_{-j})]_q|\geq a_n$ and $[\bPsi(\bX_{-j})]_q\geq b_n$, respectively, for two sequences of positive thresholds. As will be shown in Section~\ref{sec:theory}, this hard-thresholding procedure consistently recovers the graph $G$ under slightly stronger assumptions than those required for general folded concave penalties, while offering substantial computational advantages. The complete algorithm is summarized in Algorithm~\ref{alg:1}.

\begin{algorithm}[!t]
\caption{Regression with subsequent quantile check (RESQUE)}\label{alg:1}
\begin{algorithmic}[1]
\Require Data $\{\bX_i\}_{i=1}^{n}$, quantile $\{\tau_k\}_{k=1}^{K}$;
\Ensure Estimated topological order $\pi$ and underlying graph $\widehat{G}$;
\State Initialize $S=[p]$, $\pi=[\cdot]$, $\widehat{G}=\emptyset$;
\Repeat
\State Regress $X_j$ on $\{X_l\}_{l\in S\setminus\{j\}}$ by \eqref{eq:first-stage} and regress the residual on $\{X_l\}_{l\in S\setminus\{j\}}$ by \eqref{eq:second-stage} to obtain $\widehat{\bbeta}_j^S,\widehat{\bgamma}_{j,1}^S,\dots,\widehat{\bgamma}_{j,K}^S$ for all $j\in\mathcal{S}$;
\For{$j\in S$}
\If{$\widehat{\bgamma}_{j,1}^S=\dots=\widehat{\bgamma}_{j,K}^S$}
\State Update topological order by $\pi:=[j,\pi]$ and update $S:=S\setminus\pi$;
\State Concave penalty method: run the regressions models \eqref{eq:lla-select} with initial estimates $\widehat{\bbeta}_j^S,\widehat{\bgamma}_{j,1}^S$; update the estimated graph $\widehat{G}$ through
\begin{equation*}
    \begin{split}
        \widehat{\text{pa}}(j):=&\{i\in S:\exists q\text{ s.t. }[\widehat{\bbeta}^C_j]_q\neq 0\text{ and }[\bPhi(\{X_l\}_{l\in S\setminus\{j\}})]_q\text{ contains }X_i\}\\
        &\cup\{i\in S:\exists q\text{ s.t. }[\widehat{\bgamma}^C_{j,1}]_q\neq 0\text{ and }[\bPsi(\{X_l\}_{l\in S\setminus\{j\}})]_q\text{ contains }X_i\};
    \end{split}
\end{equation*}
\State Hard-thresholding method: let $t$ be a positive number depending on the current topological order $\pi$, and update the estimated graph $\widehat{G}$ by the following criterion,
\begin{equation*}
    \begin{split}
        \widehat{\text{pa}}(j):=&\{i\in S:\exists q\text{ s.t. }|[\widehat{\bbeta}_j^S]_q|\geq t\text{ and }[\bPhi(\{X_l\}_{l\in S\setminus\{j\}})]_q\text{ contains }X_i\}\\
        &\cup\{i\in S:\exists q\text{ s.t. }|[\widehat{\bgamma}_{j,1}]_q|\geq t\text{ and }[\bPsi(\{X_l\}_{l\in S\setminus\{j\}})]_q\text{ contains }X_i\};
    \end{split}
\end{equation*}
\EndIf
\EndFor
\Until $S=\emptyset$;

\end{algorithmic}
\end{algorithm}

\section{Theory}
\label{sec:theory}

This section studies the theoretical properties of RESQUE. The results are organized into two parts. First, we present convergence guarantees for the proposed regression estimators. Then, we establish consistency results for the estimated topological ordering and graph obtained by RESQUE.

\subsection{Convergence rates for regressions}
\label{sec:reg}

We present the theoretical properties of the proposed regression procedures. For each $j=1,\dots,p$, under model \eqref{eq:gam-s}, define the true parameter $\bbeta_j$ as
\begin{equation}
\label{eq:first_stage_true}
\bbeta_j=\arg\min_{\bbeta}\E[(X_{j}-\bPhi(\bX_{-j})^{\top}\bbeta)]^2.
\end{equation}
The corresponding approximation error is given by $r_j^{(f)}=\mathbb{E}(X_j\mid\bX_{-j})-\bPhi(\bX_{-j})^{\top}\bbeta_j$. The regression error is define as $\varepsilon_{j}^{\prime}=X_{j}-\mathbb{E}(X_j\mid\bX_{-j})$. We impose the following conditions on the covariate $\bPhi(\bX_{-j})$ for $j=1,\dots,p$ and regression error $\varepsilon_{j}^{\prime}$.

\begin{assumption}
\label{ass:mean_model}
Model \eqref{eq:gam-s} satisfies $\|\bbeta_j\|_0\leq s_1$ and $\|r_{j}^{(f)}\|_2^2\leq Cs_1/n$ for a constant $C>0$ and $j\in [p]$. The variance of $\varepsilon_{j}^{\prime}$ are bounded from above by a constant $C$ and the $k$th component of $\bPhi(\bX_{-j})\varepsilon_{j}^{\prime}$ satisfies Cram\'er's condition for $1\leq j\leq p$, $1\leq k\leq c_d\cdot p$.
\end{assumption}

Assumption \ref{ass:mean_model} assumes the regression coefficients in \eqref{eq:first_stage_true} are sparse, and the approximation errors are relatively small. Such conditions are standard in high-dimensional literature \citep{belloni2011inference, belloni2019valid, kalisch2007estimating}. %The sparsity requirement is satisfied when the maximum node degree in $G$ does not exceed $O(s_1)$. %Similar assumptions are standard in high-dimensional structure learning literature; see, for example, \cite{kalisch2007estimating} and \cite{ha2016penpc} for the PC algorithm. 
For the variance condition, when $X_j$ is a sink node, the regression error $\varepsilon_j^{\prime}$ coincides with $\exp\{\bPsi(\bX_{\pa^{G}(j)})^{\top}\bgamma^*_j+r^{(g)}_{j}\}\varepsilon_{j}$. In this case, the finite variance condition is equivalent to requiring that $\exp\{\bPsi(\bX_{i,\pa^{G}(j)})^{\top}\bgamma^*_j+r^{(g)}_{i,j}\}$ be bounded and that $\varepsilon_{j}$ have finite variance. When $X_j$ is not a sink node, the regression error has a smaller variance than the structural error due to conditioning.

\begin{assumption}
\label{ass:first_re}
    For $\forall j\in[p]$ and some constant $c_0\geq 1$, define $A_j=\{\bbeta\mid\|\bbeta_{T_j^c}\|_1\leq c_0\|\bbeta_{T_j}\|_1\}$, where $T_j$ is the support of $\bbeta_j$. The sample covariance matrix satisfies
    \begin{equation*}
        \kappa_1:=\inf_{1\leq j\leq p}\inf_{\bbeta\neq\boldsymbol{0},\bbeta\in A_j}\frac{\bbeta^{\top}\sum_{i=1}^{n}\bPhi(\bX_{i,-j})\bPhi(\bX_{i,-j})^{\top}\bbeta}{n\|\bbeta\|_2^2}>\kappa>0.
    \end{equation*}
\end{assumption}

Assumption \ref{ass:first_re} is a restricted eigenvalue condition \cite{bickel2009simultaneous}, which holds with high probability when $\bPhi(\bX)$ follows a sub-Gaussian \citep{mendelson2008uniform} or sub-exponential distribution \cite{adamczak2011restricted}. 

The following lemma shows the theoretical property of the $\widehat{\bbeta}_j$.

\begin{lemma}
\label{lem:first-stage}
Under Assumptions \ref{ass:mean_model}--\ref{ass:first_re}, for any constant $C>0$, there exist positive constants $C_1,C_2,C_3>0$, with the choice $\lambda_1=C_1\sqrt{\log\{\max(n,p)\}/n}$, the following bounds hold with probability at least $1-\exp[-C\log\{\max(n,p)\}]$ for all $1\le j\le p$:
\begin{equation}
\label{eq:first-stage_error}
    \|\widehat{\bbeta}_j-\bbeta_j\|_2\leq C_2\sqrt{s_1\log\{\max(n,p)\}/n}\text{ and }    \|\widehat{\bbeta}_j-\bbeta_j\|_1\leq C_3s_1\sqrt{\log\{\max(n,p)\}/n}.
\end{equation}
\end{lemma}
Following Lemma \ref{lem:first-stage}, we establish the theoretical properties of the second-stage quantile regression. Rather than analyzing the global minimizer of the nonconvex objective, we focus on the theoretical behavior of the estimator obtained via the LLA. Let $(\widehat{\bgamma}_{j,k}^{(0)},\widehat{b}_{j,k}^{(0)})_{k=1}^{K}$ denote the solution to the penalized composite quantile regression problem
\begin{equation*}
\begin{split}
\{(\widehat{\bgamma}_{j,k}^{(0)},\widehat{b}_{j,k}^{(0)})\}_{k=1}^{K}=\operatorname*{argmin}_{\{(\bgamma_{k},b_{k})\}_{k=1}^{K}}&\sum_{k=1}^{K}\bigg[\frac{1}{n}\sum_{i=1}^{n}\rho_{\tau_k}\{\widehat{Y}_{i,j}-\bPsi(\bX_{i,-j})^{\top}\bgamma_{k}-b_{k}\}+\lambda_2\|\bgamma_{k}\|_1\bigg],
\end{split}
\end{equation*}
where $\widehat{Y}_{i,j}=\log|X_{i,j}-\bPhi(\bX_{i,-j})^{\top}\widehat{\bbeta}_j|$. Then, the LLA estimator is defined as
\begin{equation*}
\begin{split}
\label{eq:second-stage_error}
(\widehat{\bgamma}_{j,k},\widehat{b}_{j,k})_{k=1}^{K}=\operatorname*{argmin}_{(\bgamma_{k},b_{k})_{k=1}^{K}}&\sum_{k=1}^{K}\bigg[\frac{1}{n}\sum_{i=1}^{n}\rho_{\tau_k}\{\widehat{Y}_{i,j}-\bPsi(\bX_{i,-j})^{\top}\bgamma_{k}-b_{k}\}+\lambda_2\|\bgamma_{j,k}\|_1\bigg]\\
&+\sum_{k=1}^{K}P_{\lambda_3}^{\prime}(\|\widehat{\bgamma}_{j,k}^{(0)}-\widehat{\bgamma}_{j,k+1}^{(0)}\|_2)\|\bgamma_{j,k}-\bgamma_{j,k+1}\|_2.
\end{split}
\end{equation*}

To facilitate the theoretical analysis, we also impose regularity conditions on the covariate $\bPsi(\bX_{-j})$ and the regression error $\log|\varepsilon_{j}^{\prime}|$ for $j=1,\dots,p$.

\begin{assumption}
\label{ass:quantile_model}
    The $K$ is bounded above by a fixed constant and the pre-selected quantiles satisfy $\tau_0\leq\min_{1\leq k\leq K}\{\tau_k\}\leq\max_{1\leq k\leq K}\{\tau_k\}\leq 1-\tau_0$ for a fixed constant $0<\tau_0<1$. The $\tau_k$th quantile of $\log(|\varepsilon_j^{\prime}|)$ satisfies $Q_{\tau_k}(X_j\mid\boldsymbol{X}_{-j})=\bPsi(\bX_{-j})^{\top}\bgamma_{j,k}+r_{j,k}^{(g)}+b_{j,k}$ where $\|\bgamma_{j,k}\|_0\leq s_2$ and the approximation error satisfies $\|r_{j,k}^{(g)}\|_2^2\leq Cs_2/n$ for a constant $C>0$ and $1\leq j\leq p$ and $1\leq k\leq K$. The $k$th component of $\bPsi(\bX_{-j})$ satisfying Cram\'er's condition for $1\leq j\leq p$, $1\leq k\leq c_d\cdot p$.
\end{assumption}

Assumption \ref{ass:quantile_model}, analogous to Assumption \ref{ass:mean_model}, requires that the regression coefficients are sparse and that the approximation errors remain small. 
Moreover, when $X_j$ is a sink node, the transformed error term admits the representation $\bPsi(\bX_{i,\pa^{G}(j)})^{\top}\bgamma^*_j+r^{(g)}_{i,j}+\log|\varepsilon_j|$. Under this setting, the conditional quantile regression parameters satisfy $\bgamma_{j,1}=\dots=\bgamma_{j,K}$ and the non-zero elements of $\bgamma_{j,1}$ are identical to $\bgamma_j^*$.

\begin{assumption}
\label{ass:quantile_density}
    Let the residual in the conditional quantile regression model be defined as $u_{j}:=\log|\varepsilon_j^{\prime}|-\bPsi(\bX_{-j})^{\top}\bgamma_{j,k}-b_{j,k}-r_{j,k}^{(g)}$. The conditional distribution of $u_{j}$ given $\bX_{-j}$ admits an absolutely continuous density $f_{u_j\mid \bX_{-j}}(\cdot)$, such that $\inf_{1\leq j\leq p}\inf_{1\leq k\leq K}f_{u_j\mid\bX_{-j}}(0)\geq \underline{f}$ and $\sup_{1\leq j\leq p}\sup_{1\leq k\leq K}\sup_{t\in\mathbb{R}}\{f_{u_j\mid\bX_{-j}}(t),\frac{\partial}{\partial t} f_{u_j\mid\bX_{-j}}(t)\}\leq \overline{f}$, and the density function $f_{u_j\mid\bX_{-j}}(\cdot)$ has sub-exponential tail.
\end{assumption}

Assumptions \ref{ass:quantile_model}--\ref{ass:quantile_density} collectively impose standard regularity conditions for the composite quantile regression. We assume that $K$ is finite, since identification of sink nodes can be achieved by comparing over a finite collection of quantile levels. The boundedness of the quantile levels $\tau_1,\dots,\tau_K$ is a common assumption and is essential for controlling the quantile variance. The smoothness condition is standard in the quantile regression literature \citep{zou2008composite,belloni2011quantile,he2021smoothed,tan2022high}. The tail assumption covers many commonly encountered distributions and is pivotal for our analysis due to the log transformation.

\begin{assumption}
\label{ass:second_re}
For $\forall j\in[p]$, $k\in\{1,\dots,K\}$ and some constant $c_0\geq 1$, define $A_j=\{\bgamma\mid\|\bgamma_{T_{j,k}^c}\|_1\leq c_0\|\bgamma_{T_{j,k}}\|_1\}$, where $T_{j,k}$ is the support set of $\bgamma_{j,k}$. Then
\begin{equation*}
    \kappa_2:=\inf_{1\leq j\leq p}\inf_{1\leq k\leq K}\inf_{\bgamma\neq\boldsymbol{0},\bgamma\in \textit{\textit{\textit{}}}A_j}\frac{\bgamma^{\top}\sum_{i=1}^{n}\bPsi(\bX_{i,-j})\bPsi(\bX_{i,-j})^{\top}\bgamma}{n\|\bgamma\|_2^2}\geq\kappa>0,
\end{equation*}
and
\begin{equation*}
        q:=\inf_{1\leq j\leq p}\inf_{1\leq k\leq K}\inf_{\bgamma\neq\boldsymbol{0},\bgamma\in A_j}\frac{\E[\|\bPsi(\bX_{-j})^{\top}\bgamma\|^2]^{3/2}}{\E[\|\bPsi(\bX_{-j})^{\top}\bgamma\|^3]}>0.
    \end{equation*}
\end{assumption}

The RNI condition \citep{belloni2011quantile} governs the accuracy of the quadratic approximation to the quantile regression loss over a restricted parameter set. 

We now establish the convergence rate of the proposed second-stage quantile regression estimator.

\begin{theorem}
\label{thm:quantile}
    Under the conditions in Lemma \ref{lem:first-stage} and Assumptions \ref{ass:quantile_model}--\ref{ass:second_re}, for any constant $C>0$, the following events hold with probability at least $1-\exp[-C\log\{\max(n,p)\}]$: for all $j\in[p]$ and $k\in[K]$, there exist constants $C_1,C_2,C_3>0$ such that, with the choices $\lambda_2=C_2\sqrt{s_1\log\{\max(n,p)\}/n}$ and $\lambda_3= C_3\sqrt{s_1s_2\log\{\max(n,p)\}/n}$, we have
    \begin{equation}
        \|\bgamma_{j,k}-\widehat{\bgamma}_{j,k}\|_2\leq C_1\sqrt{\frac{s_1s_2\log\{\max(n,p)\}}{n}}\quad\|\bgamma_k-\widehat{\bgamma}_k\|_1\leq C_1 s_2\sqrt{\frac{s_1\log\{\max(n,p)\}}{n}}.
    \end{equation}
    Moreover, the following structural recovery results hold:
\begin{enumerate}
    \item if $\Delta_{max}=\max_{1\leq k\leq K-1}\|\bgamma_{j,k}-\bgamma_{j,k+1}\|_{2}\geq C_4\sqrt{s_2}\lambda_2$ for a constant $C_4>0$, then
    \begin{equation*}
\widehat{\bgamma}_{j,k}\neq \widehat{\bgamma}_{j,k+1}
\quad\text{for all }k\in[K]
\text{ such that }
\|\bgamma_{j,k}-\bgamma_{j,k+1}\|_2=\Delta_{\max}.    \end{equation*}
    \item if $\bgamma_{j,1}=\dots=\bgamma_{j,K}$, then $\widehat{\bgamma}_{j,1}=\dots=\widehat{\bgamma}_{j,K}.$
\end{enumerate}
\end{theorem}

Theorem~\ref{thm:quantile} establishes two key results: convergence of the proposed second-stage estimator and consistent recovery of the structure of the quantile regression parameters. Different from the first-stage regression, the error bound in the second stage is inflated by a factor of $\sqrt{s_1}$. This is due to the second-stage response depends on estimated residuals in the first stage. Similar phenomena have been observed in two-stage estimation \citep{zhou2023cross,qiu2021inference}. When $X_j$ is a sink node, Theorem \ref{thm:quantile} implies that $X_j$ is correctly identified as a sink node according to the criterion in Section~\ref{sec:resque}. When $X_j$ is not a sink node and \textit{signal strength} satisfies $\Delta_{max}\geq C_4\sqrt{s_2}\lambda_2$, Theorem~\ref{thm:quantile} guarantees that the corresponding estimates are not shrunk to a common value. Under the decision rule in Section~\ref{sec:resque}, the node $X_j$ is therefore correctly classified as a non-sink node. %Signal strength conditions are commonly imposed in causal discovery \citep{kalisch2012causal, ha2016penpc, zhao2022learning}.

\subsection{Consistency for topological order and graph structure}
\label{sec:topo-graph}

This section establishes the theoretical properties of the estimated topological order and graph structure obtained from the proposed RESQUE procedure, as summarized in Algorithm~\ref{alg:1}. We first show the consistency of the estimated topological order.

\begin{figure}[ht]
    \centering
    \includegraphics[width=0.6\linewidth]{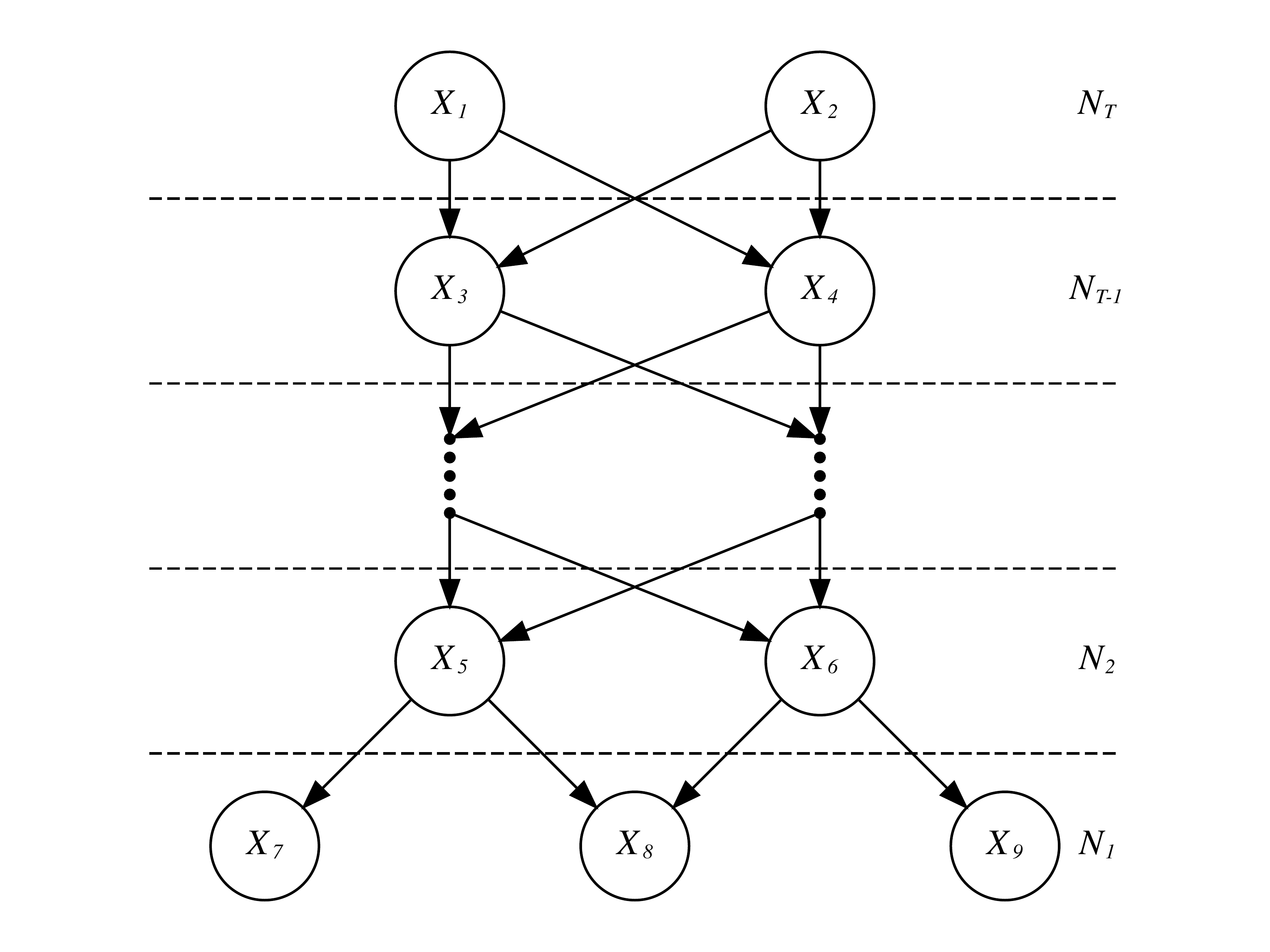}
    \caption{Illustration of the topological-layer decomposition of a DAG.}
    \label{fig:topo-order}
\end{figure}

To formalize the analysis, we introduce the notion of topological layers \citep{gao2020polynomial,zhao2022learning} of a DAG $G$. For the unknown graph $G$, a single iteration of our procedure, which involves a total of $|G|$ regression analyses, identifies all sink nodes consistently, as established in Theorem~\ref{thm:quantile}. Algorithm~\ref{alg:1} then iteratively detects sink nodes and removes them from the graph. The sink nodes identified in the first iteration constitute the first topological layer of $G$. After their removal, the sink nodes of the resulting subgraph form the second topological layer. This process continues until the graph is exhausted. Provided that each iteration correctly identifies the sink nodes, the total number of iterations required equals the maximum distance of any node from a sink node, which is referred to as the depth of the topological layering. For illustration, in Figure \ref{fig:topo-order}, the nodes $N_1=\{X_7,X_8,X_9\}$ are all sink nodes of $G$ and thus form the first topological layer. After removing $N_1$, the nodes $N_2=\{X_5,X_6\}$ become sink nodes in the reduced graph, placing them in the second layer. This procedure is repeated until the final layer $N_T=\{X_1,X_2\}$. A formal definition is given below.

Suppose that the true DAG $G$ admits a decomposition into $T$ topological layers. Define $N_1\subset[p]$ as the collection of sink nodes of $G$, corresponding to the first layer of $G$. We construct the remaining layers by iteratively partitioning the nodes as
\begin{equation*}
    N_l:=\{i\in [p]/(N_1\cup\dots\cup N_{l-1}):X_i\text{ is a sink node of graph } G/(N_1\cup\dots\cup N_{l-1})\},
\end{equation*}
where $G/(N_1\cup\dots\cup N_{l-1})$ denotes the subgraph obtained by removing the nodes $N_1\cup\dots\cup N_{l-1}$ and corresponding edges from $G$. Analogously, let $\widehat{N}_1$ be the set of sink nodes identified in the first iteration of the RESQUE algorithm. For $l\geq 2$, define
\begin{equation*}
    \widehat{N}_l:=\{i\in [p]/(\widehat{N}_1\cup\dots\cup \widehat{N}_{l-1}):\text{sink nodes identified at the }l\text{th iteration of RESQUE}\},
\end{equation*}
where $\hat{T}$ denotes the total number of iterations performed by the algorithm. The collection $\{\widehat{N}_l\}_{l=1}^{\hat{T}}$ is the estimated topological layers of the graph. For each $l$, let $\hat{G}_l=G/\{\widehat{N}_1\cup\dots\cup \widehat{N}_{l-1}\}$ denote the reduced graph and $|\hat{G}_l|$ denote the number of nodes.

We now state the assumptions required for the theoretical analysis. Assumptions~\ref{ass:first_re}–\ref{ass:second_re} are stated for the full graph $G$ and specify both the notation and conditions. When we assert that these assumptions hold for a subgraph $G\subseteq G^{\prime}$, this assertion is to be interpreted as the imposition of the same conditions after restricting all relevant notation and random variables to the index set corresponding to $G^{\prime}$. Assumption~\ref{ass:sink} generalizes the conditions in Lemma~\ref{lem:first-stage} and Theorem~\ref{thm:quantile} from a single iteration to the multi-layer setting.

\begin{assumption}[Regularty conditions on the Graph]
\label{ass:sink}
The DAG $G$ has the topological layer decomposition $G=N_1\cup\dots\cup N_T$, where the number of layers satisfies $T=O(n)$. For any $l\in\{1,\dots,T\}$, define $G_l=G/\{N_1\cup\dots\cup N_{l-1}\}$ as the subgraph obtained by removing the nodes in the first $l-1$ layers and corresponding edges. The conditions in Theorem~\ref{thm:quantile} hold for every node $j\in G_l$ and $l=\{1,\dots,T\}$.
\end{assumption}

\begin{theorem}
\label{thm:topo_order}
    Suppose that Assumption~\ref{ass:sink} holds. For each iteration $l$, let the penalty parameters be chosen as $\lambda_1=C_1\sqrt{\log(\max(|\hat{G}_l|,n))/n}$, $\lambda_2=C_2\sqrt{s_1\log(\max(|\hat{G}_l|,n))/n}$ and $\lambda_3=C_2\sqrt{s_1s_2\log(\max(|\hat{G}_l|,n))/n}$, for positive constants $C_1,C_2,C_3$. Then, with probability at least $1-\exp(-C\log(n))$, the RESQUE algorithm recovers the topological ordering, in the sense that $\hat{T}=T$ and $\widehat{N}_l=N_l$ for all $l\in[T]$.
\end{theorem}

Theorem~\ref{thm:topo_order} shows that the proposed method correctly recovers the true topological order by repeatedly applying Theorem~\ref{thm:quantile} to identify sink nodes at each iteration. In contrast to methods based on conditional independence testing after regression, our approach allows for approximate sparsity and thus provides greater flexibility. Once a causal order is obtained, regularized regression can be used to estimate the full graph. We show below that, under additional assumptions, the proposed method consistently recovers the true graph $G$.

\begin{assumption}[Signal conditions]
\label{ass:graph_strength}
 For any subgraph $G_l$ of the true underlying graph $G$ and any $j\in G_l$, the approximation errors satisfy $r_j^{(f)}=r_j^{(g)}=0$ and the true parameters in \ref{eq:gam-s} satisfy $\beta_{j,\min}:=\max_{\beta_{j,i}\neq 0}|\beta_{j,i}|\geq a\lambda_1^{\prime}$ and $\gamma_{j,\min}:=\max_{\gamma_{j,i}\neq 0}|\gamma_{j,i}|\geq a\lambda_2^{\prime}$ where the penalty parameters are given by $\lambda_1^{\prime}=C_1\sqrt{s_1\log(\max(|G_l|,n))/n}$ and $\lambda_2^{\prime}=C_2\sqrt{s_1s_2\log(\max(|G_l|,n))/n}$ for some positive constants $a,C_1,C_2$.
\end{assumption}
\begin{theorem}[Graph recovery via penalized regression]
\label{thm:graph}
    Suppose the conditions in Theorem \ref{thm:topo_order} and Assumption \ref{ass:graph_strength} hold. Then, the proposed concave penalty method consistently estimates the graph $G$ in the sense that
    \begin{equation*}
        \mathbb{P}(\hat{G}=G)\to 1\text{ as }n\to\infty.
    \end{equation*}
\end{theorem}
Theorem \ref{thm:graph} establishes that, under mild conditions on the strength of the mean and variance coefficients, the proposed method recovers the true graph $G$ with probability approaching one. Estimation of the DAG $G$ relies on selection consistency, which is guaranteed by the strong oracle optimality \citep{fan2014strong}. In high-dimensional linear regression, consistent identification of non-zero coefficients frequently requires additional conditions \citep{zhao2006model,meinshausen2009lasso}. Especially, \cite{zhao2022learning,raskutti2010restricted} employs the irrepresentability condition for the design matrix. However, as highlighted by \cite{raskutti2010restricted}, the irrepresentability condition applied to the undirected graph imposes constraints on the graph structure that are challenging to verify in practical scenarios. Therefore, our proposed method is applicable to a broader range of DAGs than the method introduced by \cite{zhao2022learning}.

The signal strength conditions in Assumption \ref{ass:graph_strength} for the mean parameters are consistent with those required for oracle estimation in high-dimensional linear regression; see \cite{fan2014strong}. The appearance of the factor $\sqrt{s_1}$ arises from converting an $\ell_2$ error bound into an $\ell_{\infty}$ bound. Under the cone invertibility factor condition of \cite{ye2010rate}, this factor can be removed by strengthening the $\ell_\infty$ error bound. Likewise, compared with the standard oracle conditions, Assumption~\ref{ass:graph_strength} imposes a stronger signal condition for the variance parameters. The additional factor $\sqrt{s_1}$ is required because the second-stage regression is based on the first-stage fit, as detailed in Theorem~\ref{thm:quantile}.

Although penalized regression methods can accurately recover the true support sets, the LLA procedure entails additional computational cost, as it requires solving both $\ell_1$ penalized linear regressions and $\ell_1$ penalized quantile regressions. In fact, estimation of the graph structure $G$ only requires identifying the supports of $\bbeta_j$ and $\bgamma_j$, which can be accomplished by selecting coefficients of sufficiently large magnitude. We consider a computationally simpler alternative to the LLA procedure based on \emph{hard thresholding} of the initial estimators $\widehat{\bbeta}_j$ and $\widehat{\bgamma}_j$. By discarding variables with small estimated coefficients, the hard-thresholding approach achieves \emph{sign consistency} in high-dimensional linear regression under slightly stronger conditions than those required for concave penalized estimation; see \cite{meinshausen2009lasso}. The following theorem shows that, under strengthened signal strength conditions, the proposed hard-thresholding procedure reconstructs the true graph $G$ with probability approaching one.

\begin{theorem}
[Graph recovery via hard-thresholding]
\label{thm:graph1}
Suppose that the conditions of Theorem~\ref{thm:graph} hold. For any subgraph $G_l$ and $j\in G_l$, assume that the true parameters in \eqref{eq:gam-s} satisfy $\beta_{j,min}:=\max_{\beta_{j,i}\neq 0}|\beta_{j,i}|\gg t_n\lambda_1^{\prime}$ and $\gamma_{j,min}:=\max_{\gamma_{j,i}\neq 0}|\gamma_{j,i}|\gg t_n\lambda_2^{\prime}$ for a positive sequence $t_n$, where $\lambda_1^{\prime}$ and $\lambda_2^{\prime}$ are defined as in Assumption~\ref{ass:graph_strength}. Then the proposed hard-thresholding procedure consistently recovers the true graph $G$ in the sense that
    \begin{equation*}
        \mathbb{P}(\hat{G}=G)\to 1\text{ as }n\to\infty.
    \end{equation*}
\end{theorem}

\section{Empirical Studies}
\label{sec:numerical-study}

This section evaluates the finite-sample performance of RESQUE across various settings.
Section \ref{subsec:sim} conducts simulation studies, and Sections \ref{subsec:realdata-sachs}--\ref{subsec:realdata-pairs} examine the proposed method using two real data benchmarks.

\subsection{Synthetic Experiments}
\label{subsec:sim}

In simulation studies, we investigate whether RESQUE can accurately recover the underlying causal graph and compare its performance with other popular causal discovery methods. 

\paragraph{Competing methods.}
We compare with other popular causal structure learning algorithms, including causal additive models (CAM) \citep{peter2014cam}, which uses high-dimensional sparse regression technique and is specifically designed for high-dimensional Gaussian noise models, rank-PC \citep{harris2013pc} uses Pearson correlation in PC algorithm \citep{kalisch2007estimating} to improve the performance for non-Gaussian data, TL \citep{zhao2022learning} which improves the RESIT algorithm \citep{peters2014causal} using sparse high-dimensional regression to detect the directed edges when determining the topological order, and non-combinatorial optimization via trace exponential and augmented lagrangian for structure learning \citep[NOTEARS]{zheng2018dags,bello2022dagma}. We utilize the R package \texttt{CAM} with default parameters to implement CAM. To implement rank-PC, we first used the R package \texttt{pcalg} \citep{kalisch2007estimating} to obtain a complete partially directed acyclic graph (CPDAG). Then, we employed the build-in function \texttt{pdag2dag} \citep{yuan2019constrained} in \texttt{pcalg} to derive a DAG. We follow the method in \cite{zhao2022learning} using the graphical Lasso \citep{friedman2008sparse} algorithm with a fixed regularization parameter and the independence test based on distance covariance measure. The significance level of independent tests in rank-PC and TL is set to $0.01$ as used by \cite{zhao2022learning}. We implement NOTEARS by the Python package \texttt{notears} \citep{zheng2018dags,bello2022dagma}. 

\paragraph{Data settings.}
The data are generated from the following structural equation model: 
\begin{equation*}
    X_j=\bX_{\pa^{G}(j)}^{\top}\bbeta_j+\exp(\bX_{\pa^{G}(j)}^{\top}\bgamma_j)\varepsilon_j\quad\text{for}\quad j=1,\dots,p.
\end{equation*}
The adjacency matrix $\mathbf U\in\{0,1\}^{p\times p}$ denotes the underlying DAG where $U_{j,k}=1$ if $j\in \text{PA}^G(k)$ and $U_{j,k}=0$ otherwise. The DAG is generated using one of three mechanisms:
\begin{itemize}
    \item {\it Random graph (random).} We consider a graph where edges are added independently with equal probability. Specifically, the adjacency matrix $\bU$ is generated at random according to the following rule: $\Pr(U_{jk}=1)=s$ if $j < k$, and $\Pr(U_{jk}=1)=0$ otherwise. Here, the parameter $s$ controls the total number of edges present in the DAG. In the simulation, we choose $s=1/p$ as in \cite{li2023nonlinear}.
    \item {\it Complete tree graph (TR).} We are focused on the $k$-ary complete tree graph, a specialized type of tree graph, where all vertices except for certain leaves have exact $k$ children. To start, we create a random $k$-ary complete tree graph, where each vertex is a simple random sample without replacement of $p$ nodes. The adjacency matrix is then determined based on the generated tree graph, with $U_{jl}=1$ if node $j$ is a child of node $l$. The above-generation process is a modification of the procedure discussed by \cite{zhao2022learning}. In our simulation, we use $k=2$ for the low-dimension setting and $k=3$ for the high-dimension setting.

    \item {\it Scale-free graph (SF).} The graph is also known as \textit{Barabási–Albert} graph. To create a new graph, we iteratively add a new node and then select $k$ existing vertices from the graph with the probability proportional to the number of existing edges. The selected nodes are connected to the added node. The generated graph is then processed by removing cycles to obtain a DAG. In the simulation, we choose $k=1$ as in \cite{zheng2018dags}.
\end{itemize}
Notice that CAM, TL, and NOTEARS are not designed for detecting the structure via the error variance component. For a comprehensive comparison, we consider the weak heterogeneity settings for parameters $\{\bbeta_j,\bgamma_j\}_{j=1}^{p}$ Each component of the mean coefficients $\bbeta_j$ is generated independently and identically from $\text{Unif}([-1,-0.5]\cup[0.5,1])$. Each component of the variance coefficients $\bgamma$ is generated independently and identically from $\text{Unif}([-1/4,-1/6]\cup[1/4,1/6])$. In the weak heterogeneity setting, the structure information is partially available through the conditional variance. For error terms, we consider three different settings:
\begin{itemize}
    \item {\it Uniform:} The random error $\varepsilon_j$ is generated independently and identically from the uniform distribution from $[-1,1]$.
    \item {\it Beta:} The random error $\varepsilon_j$ is generated independently and identically from the centered Beta distribution with shape parameters $(2,3)$.
    \item {\it Gaussian:} The random error $\varepsilon_j$ is generated independently and identically from the Gaussian distribution with mean $0$ and variance $1$.
\end{itemize}
The Gaussian distribution is often used as a standard for random errors in simulating DAG. The uniform distribution is a bounded symmetric probability distribution that remains bounded after taking the exponential. Compared to the uniform distribution, the Beta distribution has non-zero skewness.

Throughout the simulations, we set $n=200$ and consider low-dimension settings when $p\in \{5,10,15,20\}$ and high-dimension settings when $p\in \{50,75,100\}$.

\paragraph{Metrics.}
We evaluate the numerical performance based on four graph metrics: false discovery rate (FDR), false positive rate (FPR), true positive rate (TPR), and structural Hamming distance (SHD). 
To compute the metrics, let TP, RE, and FP be the numbers of identified edges with correct directions, those with wrong directions, and estimated edges not in the skeleton of the true graph. Moreover, denote by PE the total number of estimated edges, TN the number of correctly identified non-edges, and FN the number of missing edges compared to the true skeleton. Then
\begin{align*}
\textrm{FDR} &= (\textrm{RE} + \textrm{FP})/\textrm{PE}, & 
\textrm{FPR} &= (\textrm{RE} + \textrm{FP})/(\textrm{FP} + \textrm{TN}), \\
\textrm{TPR} &= {\textrm{TP}}/ 
              {(\textrm{TP} + \textrm{FN})}, &
\textrm{SHD} &= \textrm{FP} + \textrm{FN} + \textrm{RE}.
\end{align*}

\begin{figure}[ht]
    \centering
    \subfigure[FDR]{\includegraphics[width=0.49\textwidth]{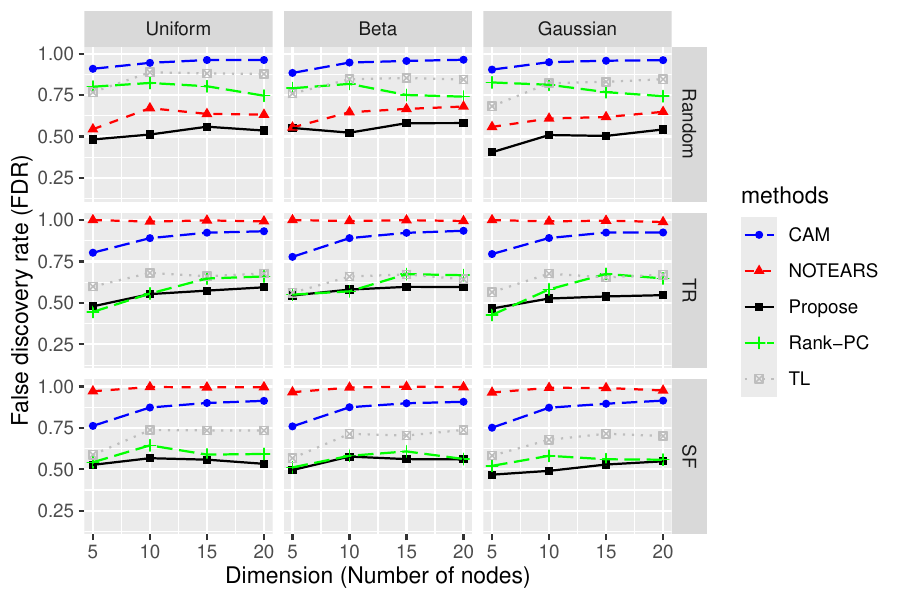}} 
    \subfigure[TPR]{\includegraphics[width=0.49\textwidth]{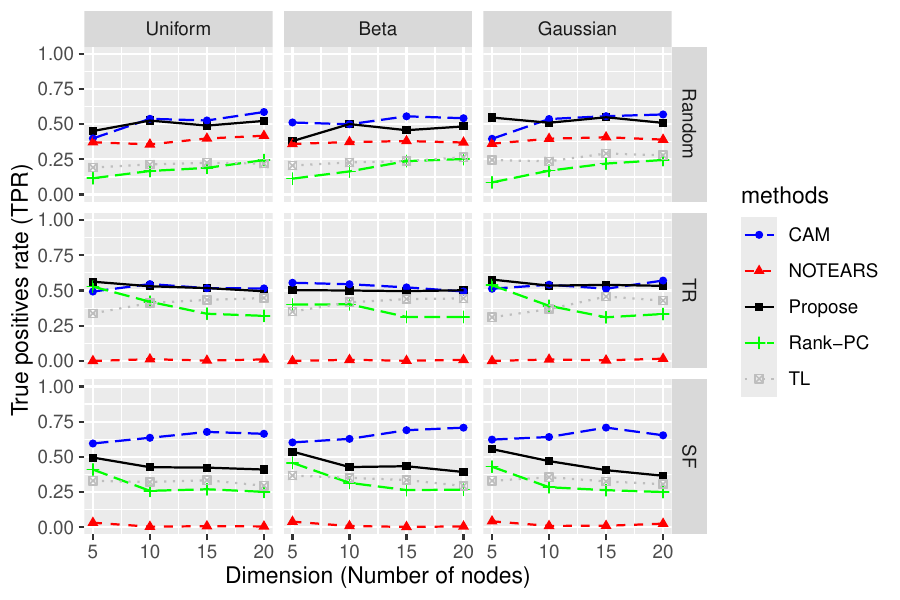}} 
    \caption{Empirical performance of CAM, rank-PC, TL, NOTEARS, and the proposed method in low-dimensional settings. False discovery rates and true positive rates in various graph structures and error distributions are presented.}
    \label{fig:ld}
\end{figure}

\begin{figure}[ht]
    \centering
    \subfigure[FDR]{\includegraphics[width=0.49\textwidth]{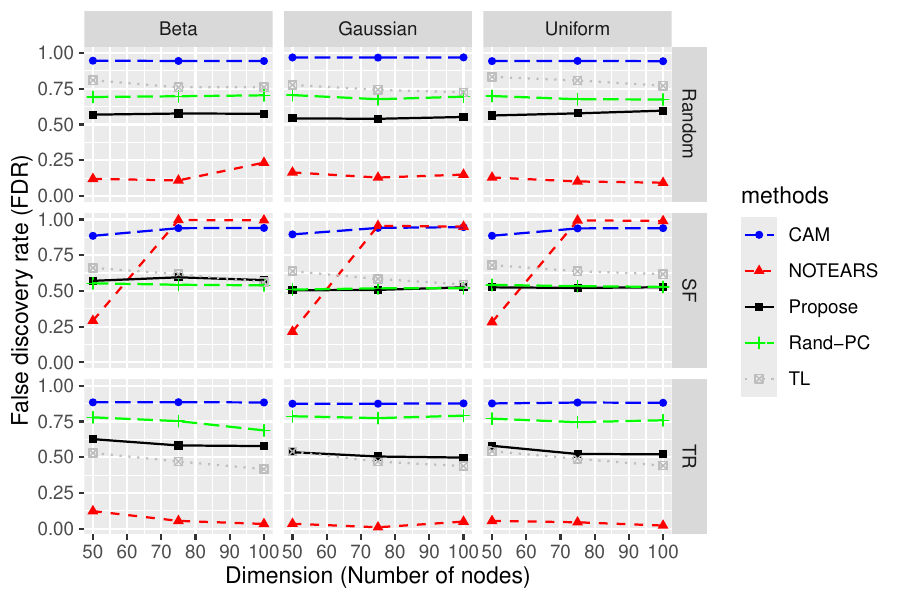}} 
    \subfigure[TPR]{\includegraphics[width=0.49\textwidth]{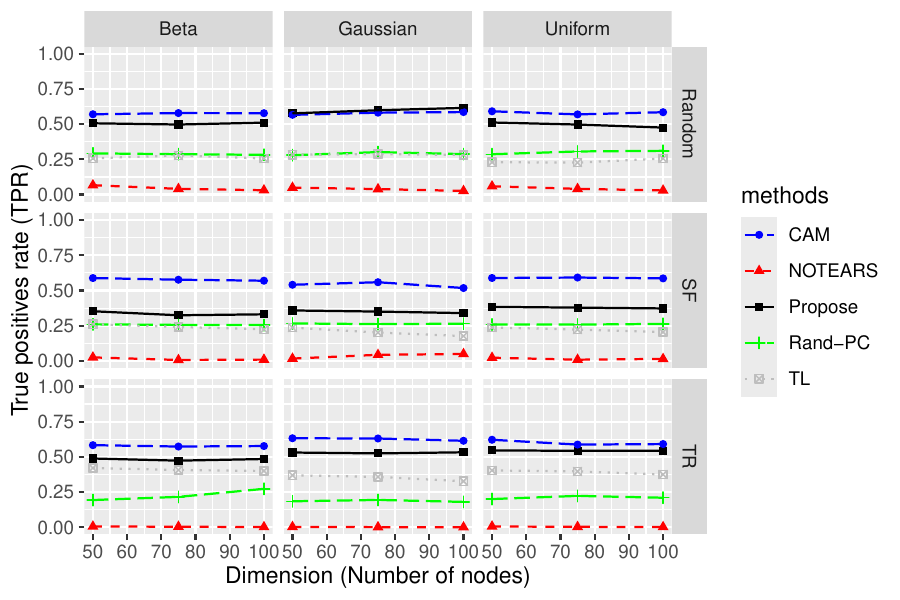}} 
    \caption{Empirical performance of CAM, rank-PC, TL, NOTEARS, and RESQUE in high-dimensional settings. False discovery rates and true positive rates in various graph structures and error distributions are presented.}
    \label{fig:hd}
\end{figure}

\paragraph{Results.}
The results for low- and high-dimensional settings are shown in Figures \ref{fig:ld}--\ref{fig:hd}, respectively. In all settings, RESQUE has high true discovery rates with relatively low false discovery rates. For example, in low-dimensional settings, it has the lowest FDR among all methods. It has nearly the highest TPR among all methods when the underlying graph is generated from Random and TR. When the underlying graph is SF, the proposed method still has a TPR higher than all other methods except CAM. Notice that CAM has a very high FDR and thus results in very high SHD, as shown in the appendix. Thus, the proposed method outperforms all other methods in the low-dimension settings. Similar results are shown in the high-dimensional settings; other than that, the NOTEARS has a very low FDR. Note that the performance of NOTEARS is sensitive to the scale of variables. After standardizing the variables, the NOTEARS shows nearly zero TPR in the high-dimensional settings. Thus, the proposed method outperforms all other methods in the high-dimension settings. We make some further comments about the SF graph. The critical requirement of our proposed method is the sparsity assumption; that is, the number of parents of every node is relatively small. A scale-free network has the property that the number of edges originating from a given node exhibits a power law distribution. Power law degree distributions alone imply that some nodes in the tail of the power law must have a high degree, which implies violating the sparsity assumption for some nodes. However, RESQUE still has reasonable performance.

\subsection{Sachs' flow cytometry dataset}
\label{subsec:realdata-sachs}

\paragraph{Dataset.}
In this section, we consider the flow cytometry dataset \cite{sachs2005causal}, which includes continuous concentrations of the molecules of numerous phosphorylated proteins and phospholipid components found in thousands of individual primary human immune system cells. The initial analysis sought to understand the causal pathways connecting a group of $p=11$ proteins using $n=7466$ observations from observational and experimental data under a series of stimulatory cues and inhibitory interventions. For evaluation, we use the \textit{consensus graph}, which is created by combining results from various experimental annotations, and is widely accepted as a reference causal graph by the biological community. 
The dataset comprises observations from nine distinct environments. As indicated in \cite{schultheiss2023ancestor}, observations from all but one environment do not follow \emph{homoscedastic linear structural equations}. %This discrepancy in validation may be attributed to the presence of non-linear effects, or heteroscedasticity, all of which align well with the capabilities of our proposed method.

\paragraph{Results.}
In Table \ref{tab:sachs}, we compare RESQUE with other methods via the number of correct estimated edges and the total structural Hamming distance. RESQUE successfully identifies 13 out of 20 directed edges in the consensus graph with the highest true positive rate among all methods and the smallest SHD. We make some further discussions of other competing methods. Coinciding with the observations in Section \ref{subsec:sim}, CAM has a relatively high TPR, detecting 12 out of 20 edges in the consensus but suffers from high false positives. The performance of Rank-PC is relatively weak compared to other recent methods, and TL and NOTEARS are based on linear structural equations and only identify edges through the conditional expectation, and thus have similar performances.

\begin{table}[ht]
    \centering
    \begin{tabular}{c c c c c c}
    \hline\hline
    &CAM&NOTEARS&Rank-PC&TL&RESQUE\\
    \hline
     \# Correct Edges & 12&7&6&6&13 \\
        SHD & 51&22&35&21&14\\
        \hline\hline
    \end{tabular}
    \caption{Comparison results of the Sachs dataset.}
    \label{tab:sachs}
\end{table}

\begin{figure}[ht]
    \centering
    \subfigure[Consensus graph  \cite{sachs2005causal}]{\includegraphics[width=0.42\textwidth]{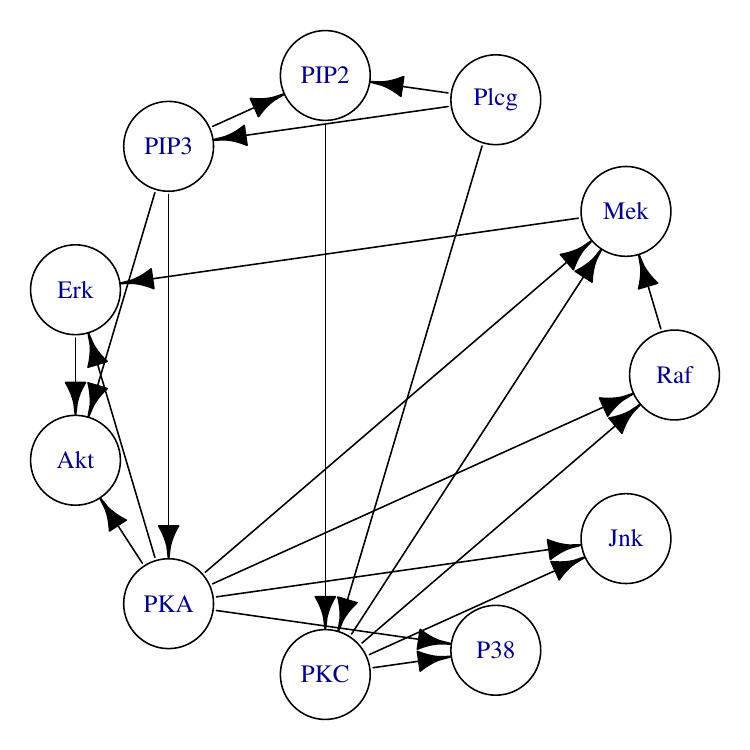}} 
    \subfigure[Estimated graph by RESQUE]{\includegraphics[width=0.49\textwidth]{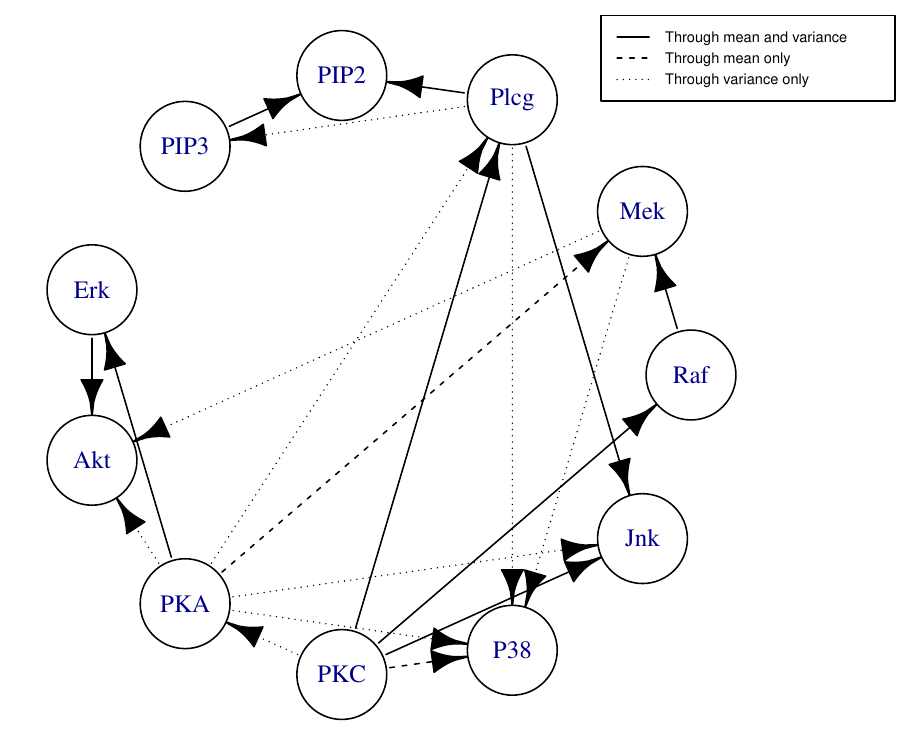}} 
    \caption{Consensus graph and estimated graph by RESQUE using the Sachs dataset.}
    \label{fig:sachs}
\end{figure}

Figure \ref{fig:sachs} displays a detailed comparison of the estimated causal graph by RESQUE and the consensus graph. %Compared to TL and NOTEARS, our proposed method identifies the directed edge through error variance components in the structural equation model. 
It shows that 4 out of 13 correctly identified edges are only identified through the variance component. In comparison, 2 out of 13 correctly identified edges are only identified through the mean component. The result suggests that the observations in the Sachs' dataset have a nontrivial heteroscedasticity, agreeing with \cite{schultheiss2023ancestor}. RESQUE identifies 9 out of 20 edges through the mean component. %The comparison suggests the failure of the linearity assumption in the estimating equation model, which agrees with the findings in \cite{schultheiss2023ancestor}. 
Note that the actual causal relations for Sachs' data have yet to be discovered. Some directed edges identified by our proposed method, while not included in the consensus graph, have also been reported by other methods: For instance, the directed edges from PKC to PKA and PKC to PKA are reported by \cite{mooij2013cyclic}. Overall, RESQUE performs well on the Sachs dataset.

\subsection{CauseEffectPairs collection}
\label{subsec:realdata-pairs}

\paragraph{Dataset.} This section considers the CauseEffectPairs collection \cite{mooij2016distinguishing}, which consists of 108 datasets, each containing samples of a pair of random variables, where one variable is known to cause the other. The goal is to determine, for each pair, which variable is the cause and which is the effect, based solely on the observed samples. Each dataset in the CauseEffectPairs collection meets the following criteria: it contains at least a few hundred samples per pair, a significant causal relation exists between the two variables, and the direction of the causal relation is known. Some pairs contain high-dimensional variables. We thus selected 99 pairs out of 108 for further analysis, where each variable in the pair is univariate. See \cite{mooij2016distinguishing} for further details.

\begin{table}[ht]
    \centering
    \begin{tabular}{c c c c c c c}
    \hline\hline
     &CAM&Rank-PC&NOTEARS&TL&RESQUE\\
    \hline
     Accuracy & 0.567&0.258&0.258&0.361&0.711 \\
        \hline\hline
    \end{tabular}
    \caption{Accuracy of CAM,
NOTEARS, DAGMA, Rank-PC, and the proposed method evaluated on CauseEffectPairs collection.}
    \label{tab:pair}
\end{table}

\paragraph{Results.}
We evaluate the performance by comparing the accuracy of causal discovery, defined as the proportion of the corrected estimated causal directions. 
% The results are presented in Table \ref{tab:pair}. 
%We also compare a new method using deep neural networks: DAGMA \citep{bello2022dagma}, a nonparametric version of NOTEARS. 
As shown in Table \ref{tab:pair}, RESQUE performs well compared to other competing methods. We also note that  CAM has relatively good performance, and the weak performances of TL and NOTEARS are partly due to the violations of their model assumptions: As shown by \cite{mooij2016distinguishing}, the scatter plots of many pairs exhibits nonlinearity and heteroscedasticity. %CAM and our proposed method all consider nonlinearity. However, our proposed method is the only one considering heteroscedasticity and thus outperforms all other methods.

\section{Discussion}
\label{sec:discussion}

This paper studies causal discovery under additive heteroscedastic structural equation models. We have established comprehensive identifiability results for a location-scale noise causal model and propose RESQUE, a two-stage regression strategy to recover the topological order and estimate the underlying DAG. Our results highlight that heteroscedasticity encodes causal direction beyond what is identifiable from the conditional mean alone, and can be systematically leveraged for causal structure learning. 

We discuss several directions that are worth pursuing. 
First, \cite{zhang2009identifiability} fully characterized non-identifiable cases of post-nonlinear causal models. It remains open and would be interesting to investigate whether similar results can also be obtained for heteroscedastic SEMs. Second, the proposed method assumes causal sufficiency, whereas many applications involve latent confounding and correlated errors \citep{li2023nonlinear}; extending the proposed framework to such settings would be a natural direction. 
%Second, many real-world systems exhibit temporal dynamics \citep{runge2019inferring}. Adapting the method to longitudinal data where both the causal structure and the heteroscedastic patterns may evolve over time is promising for future research. 
Finally, inference after causal discovery remains challenging \cite{li2020likelihood}: constructing valid confidence sets for discovered edges/orders \citep{wang2025confidence}, or for causal effects derived from an estimated graph \citep{gradu2025valid, chang2026post}, requires careful accounting for the selection inherent in the discovery process.

\section*{Acknowledgment}

This research did not receive any specific grant from funding agencies in public, commercial, or not-for-profit sectors. The work was carried out as part of the authors' routine academic duties at their respective institutions, in the absence of any commercial, financial, or institutional relationships that could be construed as a potential conflict of interest.

\spacingset{1}
\bibliographystyle{abbrv}
\bibliography{reference}

@inproceedings{lin2025skewness,
  title={A skewness-based criterion for addressing heteroscedastic noise in causal discovery},
  author={Lin, Yingyu and Huang, Yuxing and Liu, Wenqin and Deng, Haoran and Ng, Ignavier and Zhang, Kun and Gong, Mingming and Ma, Yian and Huang, Biwei},
  booktitle={International Conference on Learning Representations},
  volume={2025},
  pages={89283--89310},
  year={2025}
}

@article{park2020identifiability,
  title={Identifiability of additive noise models using conditional variances},
  author={Park, Gunwoong},
  journal={Journal of Machine Learning Research},
  volume={21},
  number={75},
  pages={1--34},
  year={2020}
}

@inproceedings{blobaum2018cause,
  title={Cause-effect inference by comparing regression errors},
  author={Bl{\"o}baum, Patrick and Janzing, Dominik and Washio, Takashi and Shimizu, Shohei and Sch{\"o}lkopf, Bernhard},
  booktitle={International Conference on Artificial Intelligence and Statistics},
  pages={900--909},
  year={2018},
  organization={PMLR}
}

@inproceedings{peters2010identifying,
  title={Identifying cause and effect on discrete data using additive noise models},
  author={Peters, Jonas and Janzing, Dominik and Sch{\"o}lkopf, Bernhard},
  booktitle={Proceedings of the thirteenth international conference on artificial intelligence and statistics},
  pages={597--604},
  year={2010},
  organization={JMLR Workshop and Conference Proceedings}
}

@inproceedings{zhang2009identifiability,
  title={On the identifiability of the post-nonlinear causal model},
  author={Zhang, Kun and Hyv{\"a}rinen, Aapo},
  booktitle={Proceedings of the Twenty-Fifth Conference on Uncertainty in Artificial Intelligence},
  pages={647--655},
  year={2009}
}

@inproceedings{zhang2010distinguishing,
  title={Distinguishing causes from effects using nonlinear acyclic causal models},
  author={Zhang, Kun and Hyv{\"a}rinen, Aapo},
  booktitle={Causality: Objectives and Assessment},
  pages={157--164},
  year={2010},
  organization={PMLR}
}

@inproceedings{hoyer2009nonlinear,
  title={Nonlinear causal discovery with additive noise models},
  author={Hoyer, PO and Janzing, D and Mooij, JM and Peters, J and Sch{\"o}lkopf, B},
  booktitle={Twenty-Second Annual Conference on Neural Information Processing Systems (NIPS 2008)},
  pages={689--696},
  year={2009},
  organization={Curran}
}

@inproceedings{spirtes2001anytime,
  title={An anytime algorithm for causal inference},
  author={Spirtes, Peter},
  booktitle={International Workshop on Artificial Intelligence and Statistics},
  pages={278--285},
  year={2001},
  organization={PMLR}
}

@article{spirtes1991algorithm,
  title={An algorithm for fast recovery of sparse causal graphs},
  author={Spirtes, Peter and Glymour, Clark},
  journal={Social science computer review},
  volume={9},
  number={1},
  pages={62--72},
  year={1991},
  publisher={Sage Publications Sage CA: Thousand Oaks, CA}
}

@article{tsamardinos2006max,
  title={The max-min hill-climbing Bayesian network structure learning algorithm},
  author={Tsamardinos, Ioannis and Brown, Laura E and Aliferis, Constantin F},
  journal={Machine learning},
  volume={65},
  number={1},
  pages={31--78},
  year={2006},
  publisher={Springer}
}

@article{vowels2022d,
  title={D’ya like {DAGs}? A survey on structure learning and causal discovery},
  author={Vowels, Matthew J and Camgoz, Necati Cihan and Bowden, Richard},
  journal={ACM Computing Surveys},
  volume={55},
  number={4},
  pages={1--36},
  year={2022},
  publisher={ACM New York, NY}
}

@article{heinze2018causal,
  title={Causal structure learning},
  author={Heinze-Deml, Christina and Maathuis, Marloes H and Meinshausen, Nicolai},
  journal={Annual Review of Statistics and Its Application},
  volume={5},
  pages={371--391},
  year={2018},
  publisher={Annual Reviews}
}

@inproceedings{xu2022inferring,
  title={Inferring cause and effect in the presence of heteroscedastic noise},
  author={Xu, Sascha and Mian, Osman A and Marx, Alexander and Vreeken, Jilles},
  booktitle={International Conference on Machine Learning},
  pages={24615--24630},
  year={2022},
  organization={PMLR}
}

@inproceedings{tran2024robust,
  title={Robust estimation of causal heteroscedastic noise models},
  author={Tran, Quang-Duy and Duong, Bao and Nguyen, Phuoc and Nguyen, Thin},
  booktitle={Proceedings of the 2024 SIAM International Conference on Data Mining (SDM)},
  pages={788--796},
  year={2024},
  organization={SIAM}
}

@inproceedings{yin2024effective,
  title={Effective causal discovery under identifiable heteroscedastic noise model},
  author={Yin, Naiyu and Gao, Tian and Yu, Yue and Ji, Qiang},
  booktitle={Proceedings of the AAAI Conference on Artificial Intelligence},
  volume={38},
  pages={16486--16494},
  year={2024}
}

@article{zhang2023statistical,
  title={Statistical insights into {HSIC} in high dimensions},
  author={Zhang, Tao and Zhang, Yaowu and Zhou, Tingyou},
  journal={Advances in Neural Information Processing Systems},
  volume={36},
  pages={19145--19156},
  year={2023}
}

@inproceedings{berrevoets2025differentiable,
  title={Differentiable causal structure learning with identifiability by notime},
  author={Berrevoets, Jeroen and Raymaekers, Jakob and Van der Schaar, Mihaela and Verdonck, Tim and Yao, Ruicong},
  booktitle={Proceedings of machine learning research},
  volume={258},
  pages={3115--3123},
  year={2025},
  organization={PMLR}
}

@article{schultheiss2023pitfalls,
  title={On the pitfalls of Gaussian likelihood scoring for causal discovery},
  author={Schultheiss, Christoph and B{\"u}hlmann, Peter},
  journal={Journal of Causal Inference},
  volume={11},
  number={1},
  pages={20220068},
  year={2023},
  publisher={De Gruyter}
}

@article{sun2023cause,
  title={Cause-effect inference in location-scale noise models: Maximum likelihood vs. independence testing},
  author={Sun, Xiangyu and Schulte, Oliver},
  journal={Advances in Neural Information Processing Systems},
  volume={36},
  pages={5447--5483},
  year={2023}
}

@article{strobl2023identifying,
  title={Identifying patient-specific root causes with the heteroscedastic noise model},
  author={Strobl, Eric V and Lasko, Thomas A},
  journal={Journal of Computational Science},
  volume={72},
  pages={102099},
  year={2023},
  publisher={Elsevier}
}

@article{shimizu2006linear,
  title={A linear {non-Gaussian} acyclic model for causal discovery},
  author={Shimizu, Shohei and Hoyer, Patrik O and Hyv{\"a}rinen, Aapo and Kerminen, Antti and Jordan, Michael},
  journal={Journal of Machine Learning Research},
  volume={7},
  number={10},
  year={2006}
}

@article{wold1954causality,
  title={Causality and econometrics},
  author={Wold, Herman},
  journal={Econometrica: Journal of the Econometric Society},
  pages={162--177},
  year={1954},
  publisher={JSTOR}
}

@incollection{bentler1988causal,
  title={Causal modeling via structural equation systems},
  author={Bentler, Peter M},
  booktitle={Handbook of multivariate experimental psychology},
  pages={317--335},
  year={1988},
  publisher={Springer}
}

@article{li2020causal,
  title={Causal discovery in physical systems from videos},
  author={Li, Yunzhu and Torralba, Antonio and Anandkumar, Anima and Fox, Dieter and Garg, Animesh},
  journal={Advances in Neural Information Processing Systems},
  volume={33},
  pages={9180--9192},
  year={2020}
}

@article{yang2024hierarchical,
  title={A hierarchical ensemble causal structure learning approach for wafer manufacturing},
  author={Yang, Yu and Bom, Sthitie and Shen, Xiaotong},
  journal={Journal of Intelligent Manufacturing},
  volume={35},
  number={6},
  pages={2961--2978},
  year={2024},
  publisher={Springer}
}

@article{wang2025confidence,
  title={Confidence sets for causal orderings},
  author={Wang, Y Samuel and Kolar, Mladen and Drton, Mathias},
  journal={Journal of the American Statistical Association},
  pages={1--14},
  year={2025},
  publisher={Taylor \& Francis}
}

@article{chang2026post,
  title={Post-selection inference for causal effects after causal discovery},
  author={Chang, Ting-Hsuan and Guo, Zijian and Malinsky, Daniel},
  journal={Biometrika},
  volume={113},
  number={1},
  pages={asaf073},
  year={2026},
  publisher={Oxford University Press}
}

@article{glymour2019review,
  title={Review of causal discovery methods based on graphical models},
  author={Glymour, Clark and Zhang, Kun and Spirtes, Peter},
  journal={Frontiers in genetics},
  volume={10},
  pages={524},
  year={2019},
  publisher={Frontiers Media SA}
}

@article{sachs2005causal,
  title={Causal protein-signaling networks derived from multiparameter single-cell data},
  author={Sachs, Karen and Perez, Omar and Pe'er, Dana and Lauffenburger, Douglas A and Nolan, Garry P},
  journal={Science},
  volume={308},
  number={5721},
  pages={523--529},
  year={2005},
  publisher={American Association for the Advancement of Science}
}

@book{spirtes2000causation,
  title={Causation, prediction, and search},
  author={Spirtes, Peter and Glymour, Clark N and Scheines, Richard},
  year={2000},
  publisher={MIT press}
}

@book{pearl2009causality,
  title={Causality},
  author={Pearl, Judea},
  year={2009},
  publisher={Cambridge university press}
}

@article{belloni2011quantile,
author = {Alexandre Belloni and Victor Chernozhukov},
title = {{l1-penalized quantile regression in high-dimensional sparse models}},
volume = {39},
journal = {The Annals of Statistics},
number = {1},
publisher = {Institute of Mathematical Statistics},
pages = {82 -- 130},
year = {2011},
doi = {10.1214/10-AOS827},
URL = {https://doi.org/10.1214/10-AOS827}
}

@article{fan2001variable,
  title={Variable selection via nonconcave penalized likelihood and its oracle properties},
  author={Fan, Jianqing and Li, Runze},
  journal={Journal of the American statistical Association},
  volume={96},
  number={456},
  pages={1348--1360},
  year={2001},
  publisher={Taylor \& Francis}
}

@article{zhang2010nearly,
  title={Nearly unbiased variable selection under minimax concave penalty},
  author={Zhang, Cun-Hui},
  journal={Annals of statistics},
  volume={38},
  number={2},
  pages={894--942},
  year={2010}
}

@article{bickel2009simultaneous,
  title={Simultaneous analysis of Lasso and Dantzig selector},
  author={Bickel, Peter J and Ritov, Ya’acov and Tsybakov, Alexandre B},
  journal={The Annals of Statistics},
  volume={37},
  number={4},
  pages={1705--1732},
  year={2009}
}

@article{raskutti2010restricted,
  title={Restricted eigenvalue properties for correlated Gaussian designs},
  author={Raskutti, Garvesh and Wainwright, Martin J and Yu, Bin},
  journal={The Journal of Machine Learning Research},
  volume={11},
  pages={2241--2259},
  year={2010},
  publisher={JMLR. org}
}

@article{mendelson2008uniform,
  title={Uniform uncertainty principle for Bernoulli and subgaussian ensembles},
  author={Mendelson, Shahar and Pajor, Alain and Tomczak-Jaegermann, Nicole},
  journal={Constructive Approximation},
  volume={28},
  pages={277--289},
  year={2008},
  publisher={Springer}
}

@article{adamczak2011restricted,
  title={Restricted isometry property of matrices with independent columns and neighborly polytopes by random sampling},
  author={Adamczak, Radoslaw and Litvak, Alexander E and Pajor, Alain and Tomczak-Jaegermann, Nicole},
  journal={Constructive Approximation},
  volume={34},
  pages={61--88},
  year={2011},
  publisher={Springer}
}

@article{tan2022high,
  title={High-dimensional quantile regression: Convolution smoothing and concave regularization},
  author={Tan, Kean Ming and Wang, Lan and Zhou, Wen-Xin},
  journal={Journal of the Royal Statistical Society Series B: Statistical Methodology},
  volume={84},
  number={1},
  pages={205--233},
  year={2022},
  publisher={Oxford University Press}
}

@article{he2021smoothed,
  title={Smoothed quantile regression with large-scale inference},
  author={He, Xuming and Pan, Xiaoou and Tan, Kean Ming and Zhou, Wen-Xin},
  journal={Journal of Econometrics},
  year={2021},
  publisher={Elsevier}
}

@article{zou2008composite,
  title={Composite quantile regression and the oracle model selection theory},
  author={Zou, Hui and Yuan, Ming},
  journal={Annals of Statistics},
  volume={36},
  number={3},
  pages={1108--1126},
  year={2008},
  publisher={Institute of Mathematical Statistics}
}

@article{li2020likelihood,
  title={Likelihood ratio tests for a large directed acyclic graph},
  author={Li, Chunlin and Shen, Xiaotong and Pan, Wei},
  journal={Journal of the American Statistical Association},
  year={2020},
  publisher={Taylor \& Francis}
}

@article{fan2014strong,
  title={Strong oracle optimality of folded concave penalized estimation},
  author={Fan, Jianqing and Xue, Lingzhou and Zou, Hui},
  journal={Annals of statistics},
  volume={42},
  number={3},
  pages={819},
  year={2014},
  publisher={NIH Public Access}
}

@article{peter2014cam,
author = {Peter B{\"u}hlmann and Jonas Peters and Jan Ernest},
title = {{CAM: Causal additive models, high-dimensional order search and penalized regression}},
volume = {42},
journal = {The Annals of Statistics},
number = {6},
publisher = {Institute of Mathematical Statistics},
pages = {2526 -- 2556},
year = {2014},
doi = {10.1214/14-AOS1260},
URL = {https://doi.org/10.1214/14-AOS1260}
}

@article{harris2013pc,
  title={{PC} algorithm for nonparanormal graphical models.},
  author={Harris, Naftali and Drton, Mathias},
  journal={Journal of Machine Learning Research},
  volume={14},
  number={11},
  year={2013}
}

@article{kalisch2007estimating,
  title={Estimating high-dimensional directed acyclic graphs with the PC-algorithm.},
  author={Kalisch, Markus and B{\"u}hlman, Peter},
  journal={Journal of Machine Learning Research},
  volume={8},
  number={3},
  year={2007}
}

@article{zhao2022learning,
  title={Learning linear non-Gaussian directed acyclic graph with diverging number of nodes},
  author={Zhao, Ruixuan and He, Xin and Wang, Junhui},
  journal={The Journal of Machine Learning Research},
  volume={23},
  number={1},
  pages={12314--12347},
  year={2022},
  publisher={JMLRORG}
}

@article{peters2014causal,
  title={Causal Discovery with Continuous Additive Noise Models},
  author={Peters, Jonas and Mooij, Joris M and Janzing, Dominik and Sch{\"o}lkopf, Bernhard},
  journal={Journal of Machine Learning Research},
  volume={15},
  pages={2009--2053},
  year={2014}
}

@article{zheng2018dags,
  title={Dags with no tears: Continuous optimization for structure learning},
  author={Zheng, Xun and Aragam, Bryon and Ravikumar, Pradeep K and Xing, Eric P},
  journal={Advances in neural information processing systems},
  volume={31},
  year={2018}
}

@inproceedings{bello2022dagma,
   author = {Bello, Kevin and Aragam, Bryon and Ravikumar, Pradeep},
   booktitle = {Advances in Neural Information Processing Systems},
   title = {{DAGMA: Learning DAGs via M-matrices and a Log-Determinant Acyclicity Characterization}},
   year = {2022}
}

@article{yuan2019constrained,
  title={Constrained likelihood for reconstructing a directed acyclic Gaussian graph},
  author={Yuan, Yiping and Shen, Xiaotong and Pan, Wei and Wang, Zizhuo},
  journal={Biometrika},
  volume={106},
  number={1},
  pages={109--125},
  year={2019},
  publisher={Oxford University Press}
}

@article{friedman2008sparse,
  title={Sparse inverse covariance estimation with the graphical lasso},
  author={Friedman, Jerome and Hastie, Trevor and Tibshirani, Robert},
  journal={Biostatistics},
  volume={9},
  number={3},
  pages={432--441},
  year={2008},
  publisher={Oxford University Press}
}

@article{belloni2011inference,
  title={Inference for high-dimensional sparse econometric models},
  author={Belloni, Alexandre and Chernozhukov, Victor and Hansen, Christian},
  journal={arXiv preprint arXiv:1201.0220},
  year={2011}
}

@article{belloni2019valid,
  title={Valid post-selection inference in high-dimensional approximately sparse quantile regression models},
  author={Belloni, Alexandre and Chernozhukov, Victor and Kato, Kengo},
  journal={Journal of the American Statistical Association},
  volume={114},
  number={526},
  pages={749--758},
  year={2019},
  publisher={Taylor \& Francis}
}

@article{koenker1978regression,
  title={Regression quantiles},
  author={Koenker, Roger and Bassett, Jr, Gilbert},
  journal={Econometrica: journal of the Econometric Society},
  pages={33--50},
  year={1978},
  publisher={JSTOR}
}

@article{zhou2023cross,
  title={Cross-Fitted Residual Regression for High-Dimensional Heteroscedasticity Pursuit},
  author={Zhou, Le and Zou, Hui},
  journal={Journal of the American Statistical Association},
  volume={118},
  number={542},
  pages={1056--1065},
  year={2023},
  publisher={Taylor \& Francis}
}

@article{qiu2021inference,
  title={Inference of heterogeneous treatment effects using observational data with high-dimensional covariates},
  author={Qiu, Yumou and Tao, Jing and Zhou, Xiao-Hua},
  journal={Journal of the Royal Statistical Society Series B: Statistical Methodology},
  volume={83},
  number={5},
  pages={1016--1043},
  year={2021},
  publisher={Oxford University Press}
}

@article{gao2020polynomial,
  title={A polynomial-time algorithm for learning nonparametric causal graphs},
  author={Gao, Ming and Ding, Yi and Aragam, Bryon},
  journal={Advances in Neural Information Processing Systems},
  volume={33},
  pages={11599--11611},
  year={2020}
}

@article{zhao2006model,
  title={On model selection consistency of Lasso},
  author={Zhao, Peng and Yu, Bin},
  journal={The Journal of Machine Learning Research},
  volume={7},
  pages={2541--2563},
  year={2006},
  publisher={JMLR. org}
}

@article{meinshausen2009lasso,
  title={Lasso-type recovery of sparse representations for high-dimensional data},
  author={Meinshausen, Nicolai and Yu, Bin},
  journal={The Annals of Statistics},
  volume={37},
  number={1},
  pages={246--270},
  year={2009}
}

@article{ye2010rate,
  title={Rate minimaxity of the Lasso and Dantzig selector for the lq loss in lr balls},
  author={Ye, Fei and Zhang, Cun-Hui},
  journal={The Journal of Machine Learning Research},
  volume={11},
  pages={3519--3540},
  year={2010},
  publisher={JMLR. org}
}

@article{li20081,
  title={L 1-norm quantile regression},
  author={Li, Youjuan and Zhu, Ji},
  journal={Journal of Computational and Graphical Statistics},
  volume={17},
  number={1},
  pages={163--185},
  year={2008},
  publisher={Taylor \& Francis}
}

@article{schultheiss2023ancestor,
  title={Ancestor regression in linear structural equation models},
  author={Schultheiss, Christoph and B{\"u}hlmann, Peter},
  journal={Biometrika},
  volume={110},
  number={4},
  pages={1117--1124},
  year={2023},
  publisher={Oxford University Press}
}

@inproceedings{mooij2013cyclic,
  title={Cyclic causal discovery from continuous equilibrium data},
  author={Mooij, Joris M and Heskes, Tom},
  booktitle={Proceedings of the Twenty-Ninth Conference on Uncertainty in Artificial Intelligence},
  pages={431--439},
  year={2013}
}

@article{mooij2016distinguishing,
  title={Distinguishing cause from effect using observational data: methods and benchmarks},
  author={Mooij, Joris M and Peters, Jonas and Janzing, Dominik and Zscheischler, Jakob and Sch{\"o}lkopf, Bernhard},
  journal={Journal of Machine Learning Research},
  volume={17},
  number={32},
  pages={1--102},
  year={2016}
}

@article{li2023nonlinear,
  title={Nonlinear causal discovery with confounders},
  author={Li, Chunlin and Shen, Xiaotong and Pan, Wei},
  journal={Journal of the American Statistical Association},
  pages={1--10},
  year={2023},
  publisher={Taylor \& Francis}
}

@inproceedings{immer2023identifiability,
  title={On the identifiability and estimation of causal location-scale noise models},
  author={Immer, Alexander and Schultheiss, Christoph and Vogt, Julia E and Sch{\"o}lkopf, Bernhard and B{\"u}hlmann, Peter and Marx, Alexander},
  booktitle={International Conference on Machine Learning},
  pages={14316--14332},
  year={2023},
  organization={PMLR}
}

@article{scholkopf2021toward,
  title={Toward causal representation learning},
  author={Sch{\"o}lkopf, Bernhard and Locatello, Francesco and Bauer, Stefan and Ke, Nan Rosemary and Kalchbrenner, Nal and Goyal, Anirudh and Bengio, Yoshua},
  journal={Proceedings of the IEEE},
  volume={109},
  number={5},
  pages={612--634},
  year={2021},
  publisher={IEEE}
}

@article{maathuis2010predicting,
  title={Predicting causal effects in large-scale systems from observational data},
  author={Maathuis, Marloes H and Colombo, Diego and Kalisch, Markus and B{\"u}hlmann, Peter},
  journal={Nature methods},
  volume={7},
  number={4},
  pages={247--248},
  year={2010},
  publisher={Nature Publishing Group US New York}
}

@article{chickering2002optimal,
  title={Optimal structure identification with greedy search},
  author={Chickering, David Maxwell},
  journal={Journal of machine learning research},
  volume={3},
  number={Nov},
  pages={507--554},
  year={2002}
}

@article{peters2014identifiability,
  title={Identifiability of Gaussian structural equation models with equal error variances},
  author={Peters, Jonas and B{\"u}hlmann, Peter},
  journal={Biometrika},
  volume={101},
  number={1},
  pages={219--228},
  year={2014},
  publisher={Oxford University Press}
}

@article{gradu2025valid,
  title={Valid inference after causal discovery},
  author={Gradu, Paula and Zrnic, Tijana and Wang, Yixin and Jordan, Michael I},
  journal={Journal of the American Statistical Association},
  volume={120},
  number={550},
  pages={1127--1138},
  year={2025},
  publisher={Taylor \& Francis}
}

\end{document}